\documentclass[journal, final, 10pt, twocolumn]{IEEEtran}
\usepackage{graphicx}
\usepackage{amsmath}
\usepackage{subfigure}
\usepackage{epsfig}
\usepackage{float}
\usepackage{lipsum}
\usepackage{stfloats}
\usepackage{array}
\usepackage{amssymb}
\usepackage{cite}
\usepackage{url}
\usepackage{amsthm}
\usepackage{algorithm2e}
\usepackage{algorithmic}
\usepackage{multirow}
\usepackage{fancyh}
\usepackage{upgreek}
\normalsize

\newcommand{\lambdab}{\lambda}
\newcommand{\lambdad}{\uplambda}
\newcommand{\Phid}{\mathsf{\Phi}}
\newcommand{\Phic}{\Phi}
\newcommand{\Phidt}{\tilde{\mathsf{\Phi}}}
\newcommand{\lambdadt}{\tilde{\uplambda}}
\newcommand{\Pd}{\mathsf{P}}
\newcommand{\Pc}{P}

\newcommand{\SEc}{C}
\newcommand{\SEd}{\mathsf{C}}
\newcommand{\bSEc}{\bar{C}}
\newcommand{\bSEd}{\bar{\mathsf{C}}}

\newcommand{\Hco}{H_0}
\newcommand{\Hck}{H_k}
\newcommand{\bco}{s_0}
\newcommand{\bck}{s_k}
\newcommand{\yco}{y}
\newcommand{\rco}{r_0}
\newcommand{\rck}{r_k}
\newcommand{\aco}{a_0}
\newcommand{\ack}{a_k}
\newcommand{\zco}{z}

\newcommand{\Hdo}{\mathsf{H}_{0}}
\newcommand{\Hdj}{\mathsf{H}_{j}}
\newcommand{\bdo}{\mathsf{s}_{0}}
\newcommand{\bdj}{\mathsf{s}_{j}}
\newcommand{\ydo}{\mathsf{y}}
\newcommand{\rdo}{\mathsf{r}_{0}}
\newcommand{\rdj}{\mathsf{r}_{j}}
\newcommand{\ado}{\mathsf{a}_{0}}
\newcommand{\adj}{\mathsf{a}_{j}}
\newcommand{\zdo}{\mathsf{z}}

\newcommand{\etac}{\eta}
\newcommand{\etad}{{\upeta}}

\newcommand{\SIRc}{\text{SIR}}
\newcommand{\SIRd}{\mathsf{SIR}}

\newcommand{\Ic}{\sigma^2}
\newcommand{\Iinc}{\sigma^2_{\text{in}}}
\newcommand{\Ioutc}{\sigma^2_{\text{out}}}
\newcommand{\bIoutc}{\overline{\sigma^2}_{\text{out}}}

\newcommand{\Id}{\upsigma^2}
\newcommand{\Iind}{\upsigma^2_{\text{in}}}
\newcommand{\Ioutd}{\upsigma^2_{\text{out}}}
\newcommand{\bIoutd}{\overline{\upsigma^2}_{\text{out}}}

\newcommand{\rhoc}{\rho}
\newcommand{\rhod}{\varrho}

\newcommand{\erf}{\mathrm{erf}}
\newcommand{\erfc}{\mathrm{erfc}}

\newcommand{\dc}{\mathsf{d}_{\scriptscriptstyle \rm ex}}
\newcommand{\adc}{a_{\scriptscriptstyle \rm ex}}
\newcommand{\dr}{\mathsf{r}}
\newcommand{\da}{\mathsf{a}}

\newtheorem{prp}{Proposition}
\newtheorem{cor}{Corollary}
\theoremstyle{definition}

\newtheorem{exmp}{Example}

\begin{document}

\title{An Analytical Framework for Device-to-Device Communication in Cellular Networks}

\author{Geordie~George,~\IEEEmembership{Student Member,~IEEE}, Ratheesh~K.~Mungara,~\IEEEmembership{Student Member,~IEEE} and~Angel~Lozano,~\IEEEmembership{Fellow,~IEEE}
        \thanks{G. George, R. K. Mungara, and A. Lozano are with the Department of Information and Communication Technologies, Universitat Pompeu Fabra (UPF), 08018 Barcelona, Spain. E-mail: \{geordie.george, ratheesh.mungara, angel.lozano\}@upf.edu. This work was supported in part by Intel's University Research Program ``5G: Transforming the Wireless User Experience'' and by the MINECO Project TEC2012-34642. Parts of this paper were presented at the IEEE Global Communications Conference (GLOBECOM'14).}
        }

\maketitle

\begin{abstract}
This paper presents a framework that enables characterizing analytically the spectral efficiency achievable by D2D (device-to-device) communication integrated with a cellular network. This framework is based on a stochastic geometry formulation with a novel approach to the modeling of interference and with the added possibility of incorporating exclusion regions to protect cellular receivers from excessive interference from active D2D transmitters.
To illustrate the potential of the framework, a number of examples are provided. These examples confirm the potential of D2D communication in situations of strong traffic locality as well as the effectiveness of properly sized exclusion regions.
\end{abstract}


\IEEEpeerreviewmaketitle  

\section{Introduction} 
\label{sec:introduction}

Device-to-device (D2D) communication, currently being touted as a potential ingredient of 5th-generation wireless networks~\cite{Boccardi-5G,What5G,STalwar-Intel-D2D,KJohnsson-Intel-D2D}, allows users in close proximity to establish direct communication, replacing two long hops via the base station (BS) with a single shorter hop. 
Provided there is sufficient spatial locality in the wireless traffic, this can bring about several benefits: reduced power consumption, lower end-to-end latency, reduced backhaul loads, and especially a much higher spectral efficiency thanks to the shorter range and denser spectral reuse.
While the nature of D2D communication results in more complex and irregular topologies, unlike in traditional ad-hoc networks an integrated D2D system can benefit from infrastructure assistance to perform efficient user discovery, channelization, and interference management.

The 3rd Generation Partnership Project (3GPP) is in the process of studying and standardizing D2D communication for cellular networks~\cite{3GPP-TR-22-803-V100,LeiD2D,XLin-Andrews14} while problems associated with D2D have been identified and are being explored by academia~\cite{Dopp-Jan10,FodorD2D11,HynkeeD2D11,FodorD2D12,YPeiD2D13,Hao-Doppler-2013,DFengD2D13,ShalmashiD2D13,HTangD2D13,WXuD2D13,AsadiD2D14,KauffmanD2D13}.
Initially, most such works had relied on simulations.
Recognizing that stochastic geometry tools allow for models that are both amenable to analytical treatment and highly representative of the spatial behavior of D2D users, 
more recent works~\cite{Lin-Andrews-Ghos14,Ye-Andrews13,Lee-Lin-Andrews15} have modeled the user locations via PPP (Poisson point process) distributions and analytically tackled D2D communication.

In this paper, we continue down the path of \cite{Lin-Andrews-Ghos14}, but with a different approach to model interference and with a controllable degree of spatial averaging. In addition, we allow the length of the D2D links to depend on the user density,
altogether completing a powerful and flexible framework that enables characterizing in simpler form---sometimes even in closed form---the spectral efficiencies achievable with D2D communication.
The framework accommodates both \emph{underlay} or \emph{overlay} options, where respectively the D2D communication reuses the existing uplink or utilizes dedicated spectrum. 
A preliminary version of this work can be found in \cite{GeoMunLoz-D2D-GC14}, where only overlaid D2D was addressed.
We present several examples of how this framework can be leveraged to gauge the benefits of D2D, answering questions such as:
\begin{itemize}
\item How often is direct D2D better than two hops (uplink-downlink) via the BS?
\item How many D2D links can be packed on each cell without compromising the spectral efficiencies of those D2D links (bits/s/Hz) or that of the cellular user in that cell?
\item How much better is the system spectral efficiency (bits/s/Hz per cell or bits/s/Hz per unit area) given the denser spectral reuse?
\end{itemize}
It is our hope that this framework can serve other researchers as they further explore the potential and the challenges associated with D2D. 


\section{System Model}
\label{sec:sys model}

We consider an interference-limited cellular network where the BSs are regularly placed on a hexagonal grid.
(It would also be possible to model the BS locations stochastically, but in any event the emphasis in our framework is on the location of the D2D users.)
At each BS, cellular transmissions are orthogonalized while multiple D2D links share each time-frequency signaling resource.
Transmitters and receivers have a single antenna and each receiver knows the fading of only its own link, be it cellular or D2D.
Our focus is on a given time-frequency resource, where one cellular uplink and/or (underlay/overlay) multiple D2D links are active in each cell.

To facilitate the readability of the equations, we utilize distinct fonts for the cellular and D2D variables. 

\subsection{User Locations}
\label{sec:loc of TXs}

The locations of the transmitters, both cellular and D2D, are modeled relative to the location of a given receiver under consideration. 
For the \emph{cellular uplink}, the receiver under consideration is a BS whereas, for the \emph{D2D link}, it is a user.
In either case, and without loss of generality, we place such receiver at the origin and index the intended transmitter with zero. All other transmitters (interferers) are indexed in order of increasing distance within each class (cellular and D2D). 
For a cellular uplink, the intended transmitter is always the closest cellular transmitter while, for a D2D link, the intended transmitter need not be the closest D2D transmitter.

\begin{figure}  
\centering
\includegraphics[width=0.97\columnwidth]{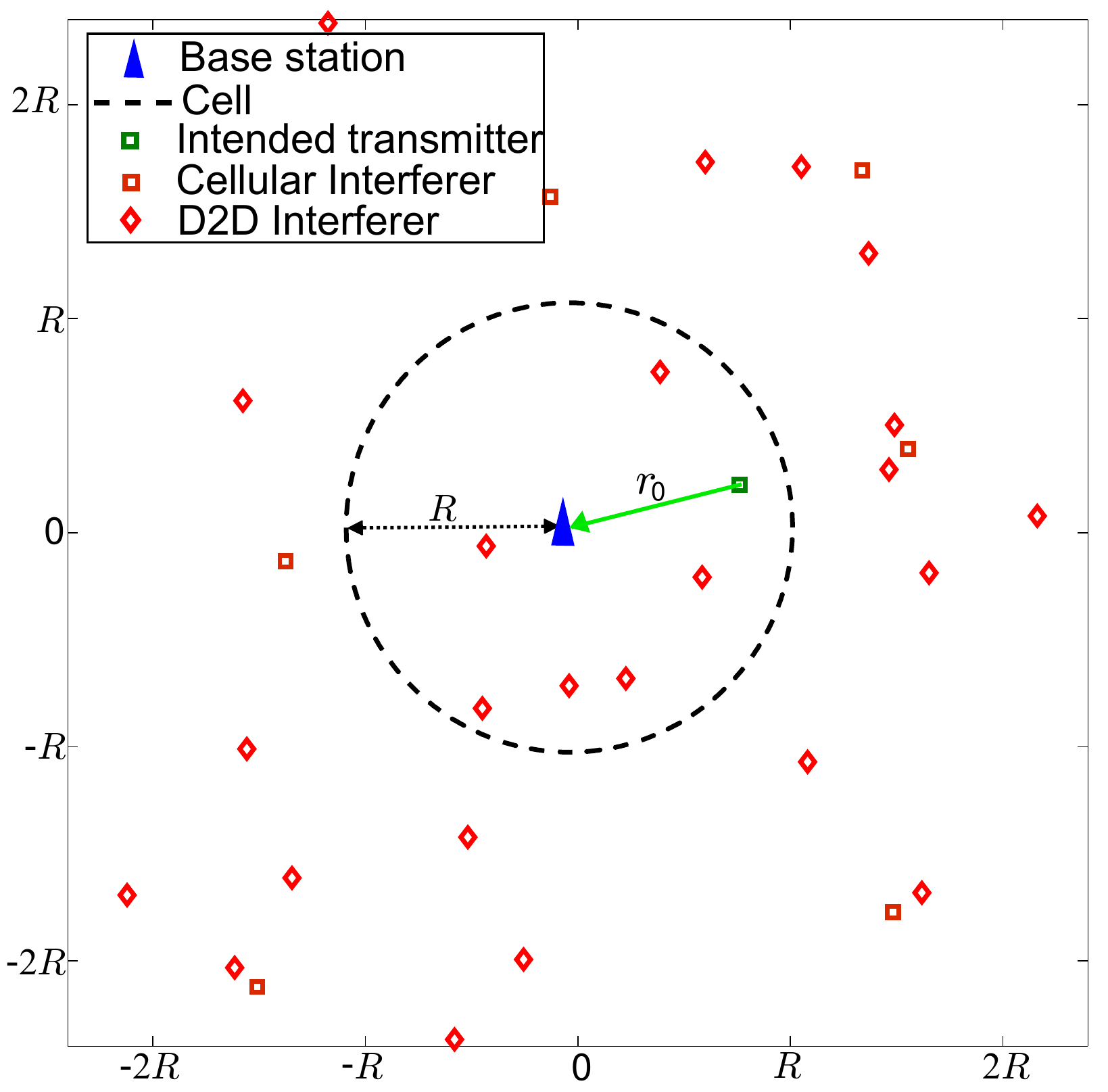} 
\vspace{-0.1in}
\centering \caption{Cellular uplink with D2D.
Located at the origin is a receiving BS and shown with a square marker in the surrounding circle is its intended cellular transmitter; shown with square markers outside the circle are the cellular interferers; shown with diamond markers are the D2D interferers.}
\label{fig:networkmodel}
\end{figure}

\subsubsection{Cellular Uplink} 
\label{sec:TXs for Uplink}


To study this link, we place a receiving BS at the origin and locate an intended cellular transmitter uniformly within the cell associated with that BS (cf. Fig.~\ref{fig:networkmodel}), which is circular with radius $R$ and denoted by $\mathcal{B}(0,R)$.
There is one and only one cellular transmitter within $\mathcal{B}(0,R)$, and its distance to the BS at the origin is denoted by $\rco$.

The cellular interferers from other cells are outside $\mathcal{B}(0,R)$, modeled via a PPP $\Phic$ with density $\lambdab = \frac{1}{\pi R^2}$. With this density made to coincide with the number of BSs per unit area, this has been shown to be a fine model for a network with one cellular transmitter per cell~\cite{Novlan-Dhillon-Andrews-COM13,Lin-Andrews-Ghos14}.

The D2D interferer locations form another independent PPP $\Phid$ with density $\lambdad = K\lambdab$ such that there are, on average, $K$ active D2D links per cell.

\subsubsection{D2D Link} 
\label{sec:TXs for D2D}

To study this link, we place a D2D receiver at the origin and locate its intended D2D transmitter at a distance $\rdo$.
Given the absence of empirical data on whether and how the length of the intended links depends on the user density, we adopt the rather general model $\rdo=\frac{\dr}{K^\upbeta}$ with $\dr>0$ and $\upbeta \geq 0$.
For strictly positive $\upbeta$, the link length shrinks as the user density intensifies---a behavior that is intuitively reasonable---whereas, for $\upbeta=0$, we obtain $\rdo = \dr$ independently of the user density.


The cellular and D2D interferers conform to $\Phic$ and $\Phid$, respectively.
Note that there may be cellular interferers arbitrarily close to a D2D receiver, a point whose implications are discussed later.

\subsection{Received Signal}
\label{sec:RX signal}
We denote by $\Pc$ and $\Pd$ the (fixed) signal powers of cellular and D2D users, respectively, both measured at $1$ m from the transmitter and with their ratio being $\mu = \Pd / \Pc$. 
Unit-gain antennas are featured at the users while the BS antenna gain is immaterial to the cellular uplink because, in interference-limited conditions, it affects signal and interference equally.

In order to present the results in the most general fashion, we define a binary parameter $\alpha \in \{0,1\}$ that distinguishes between underlay ($\alpha = 1$) and overlay ($\alpha = 0$).

\subsubsection{Cellular Uplink}

The BS at the origin observes
\begin{align} \label{eq:Signal2c}
\yco = \sqrt{\Pc \, \rco^{-\etac}} \, \Hco \, \bco + \zco
\end{align}
where the first term is the signal from the intended cellular user while the second term represents the interference
\begin{align} \label{eq:zintc}
\zco &= \sum_{k=1}^{\infty} \sqrt{\Pc \, \rck^{-\etac}} \, \Hck \, \bck + \alpha \sum_{j=1}^{\infty} \sqrt{\Pd \, \rdj^{-\etac}} \, \Hdj \, \bdj
\end{align}
whose first summation spans the other-cell cellular users in $\Phic \backslash \mathcal{B}(0,R)$ and whose second summation spans all the D2D transmitters in $\Phid$. 
In turn, $\etac > 2$ is the pathloss exponent for cellular links, $\rck$ represents the distance between the $k$th cellular transmitter and the BS at the origin, $\Hck$ denotes the corresponding fading,
and $\bck$ is the symbol transmitted by the $k$th cellular transmitter. Similarly, $\rdj$ represents the distance between the $j$th D2D transmitter and the BS at the origin, $\Hdj$ denotes the corresponding fading and $\bdj$ is the symbol transmitted by the $j$th D2D transmitter. The fading coefficients $\Hck$ and $\Hdj$ are independent identically distributed (IID) complex Gaussian random variables with zero mean and unit variance, i.e., drawn from $\mathcal{N}_{\mathbb{C}}(0,1)$. Likewise, $\bck \sim \mathcal{N}_{\mathbb{C}}(0,1)$ and $\bdj \sim \mathcal{N}_{\mathbb{C}}(0,1)$.

\subsubsection{D2D Link}
 
To analyze this link, we shift the origin to the D2D receiver of interest, which observes
\begin{align} \label{eq:Signal2d}
\ydo = \sqrt{\Pd \, \rdo^{-\etad}} \, \Hdo \, \bdo + \zdo
\end{align}
where the first term is the signal from the intended D2D transmitter while the second term is the interference
\begin{align} \label{eq:zintd}
\zdo = \alpha \sum_{k=1}^{\infty} \sqrt{\Pc \, \rck^{-\etad}} \, \Hck \, \bck + \sum_{j=1}^{\infty} \sqrt{\Pd \, \rdj^{-\etad}} \, \Hdj \, \bdj
\end{align}
received from the D2D transmitters in $\Phid$ and the cellular transmitters in $\Phic$. For these user-to-user signals, we consider a different pathloss exponent $\etad > 2$.

\section{Interference Modeling}
\label{sec:Intf model}


A key differentiating feature of our framework is the interference modeling, expounded in this section and validated later in the paper.
We depart from the approach in \cite{Lin-Andrews-Ghos14} where $\zco$ and $\zdo$ are
explicitly modeled as per (\ref{eq:zintc}) and (\ref{eq:zintd}), with all the products of signal and fading Gaussian variates therein, and also from the approach in \cite{Heath-Kount-Tian13} where a Gamma distribution with matched moments is fitted.

Rather, recognizing that both $\zco$ and $\zdo$ consist of a large number of independent terms whose fading is unknown by the receiver of interest, we model their short-term distributions as zero-mean complex Gaussian with matched conditional covariances $\Ic = \mathbb{E}\left[|\zco|^2 | \{\rck,\rdj\}\right]$ and $\Id = \mathbb{E}\left[|\zdo|^2 | \{\rck,\rdj\}\right]$, respectively, where the expectations are over the data and fading distributions.
The conditional covariance $\Ic$, which represents the power of $\zco$ for given interferer locations, is easily found to equal
\begin{align} 
\Ic &=
   \sum_{k=1}^{\infty} \Pc \, \rck^{-\etac} + \alpha \sum_{j=1}^{\infty} \Pd \, \rdj^{-\etac}
\label{eq:I_power1}
\end{align}
while its D2D counterpart $\Id$ equals
\begin{align} 
\Id 
\label{eq:I_power5b}
  &= \alpha \sum_{k=1}^{\infty}\Pc \, \rck^{-\etad} + \sum_{j=1}^{\infty} \Pd \, \rdj^{-\etad}.
\end{align}

Besides the central limit theorem, which renders $\zco$ and $\zdo$ close to Gaussian, there are information-theoretic arguments in favor of formally modeling the aggregate interference as Gaussian with a power dictated by the locations of the interferers: if the exact distribution of the interference is either unknown or ignored by the receiver, with a decoder designed to handle Gaussian noise, then the achievable spectral efficiency is precisely as if the interference were indeed Gaussian \cite{lapidoth2002fading}. 
Thus, the results obtained under our model not only approximate closely (always as lower bounds because Gaussian noise is the worst-case noise \cite{worst-case-additive-noise})
the values obtained with $\zco$ and $\zdo$ as per (\ref{eq:zintc}) and (\ref{eq:zintd}), but arguably these results are more operationally relevant because it is unlikely that the receiver can learn the distributions of $\zco$ and $\zdo$, and even if it could a standard decoder for Gaussian noise be featured.


Returning to our modeling approach, both (\ref{eq:I_power1}) and (\ref{eq:I_power5b}) contain an infinite number of terms, of which a handful largely dominate the total interference power because of the distance-dependent pathloss.
In recognition of this, we condition on the interferer locations within a circle surrounding the receiver of interest and replace the aggregate interference emanating from outside that circle with its expected (over the interference locations) value. As we shall see, this expected value is representative of most instances of the interference outside the circle---by virtue of the law of large numbers---and, thanks to the potency of stochastic geometry, this expected value can be computed explicitly.   
The introduction of the averaging circle allows reducing the number of variables retained in the formulation without the significant loss of information brought about by a complete spatial averaging, which is a typical recourse in stochastic geometry analyses. This allows establishing the performance for specific locations of the users within the circle, which are the dominant interferers, and not only the average performance over all such locations.
The radius of the averaging circle then becomes a modeling parameter that should be chosen to balance simplicity (the smaller the circle, the fewer interferers that are explicitly retained) and accuracy (the smaller the circle, the less fidelity in representing interference instances outside the circle with their average). Remarkably, an averaging circle encompassing very few interferers suffices, something that is illustrated in Example \ref{example2} and Fig. \ref{Fig:val3}: conditioning on the location of the three closest interferes suffices to capture the specificity of various situations, and the rest of the interference can be replaced by its spatial average with hardly any loss in fidelity. For each specific in-circle situation in the example, simulation results for 10 different snapshots of the out-of-circle interference are shown, and all are tightly clustered around the analytical result involving their spatial averages; in fact, for some of the situations it is utterly impossible to tell that multiple snapshots are overlapped.
With an averaging circle that encompassed more than 3 interferers, the accuracy would increase even further.

A natural and very safe choice is to have the size of the circle coincide with that of a cell, $\mathcal{B}(0,R)$.
Unless otherwise stated such is the size of the averaging circle, whereby the interference power in (\ref{eq:I_power1}) can be rewritten as
\begin{align} \label{eq:I_power2}
\Ic = \underbrace{ \alpha \sum_{j=1}^{K'} \Pd \, \rdj^{-\etac} }_{\Iinc} + \underbrace{\alpha \!\!\!\! \sum_{j = K'+1}^{\infty} \!\!\!\! \Pd \, \rdj^{-\etac} +  \sum_{k=1}^{\infty} \Pc \, \rck^{-\etac}}_{\Ioutc}
\end{align}
where $\Iinc$ corresponds to the $K'$ D2D transmitters in $\Phid \cap \mathcal{B}(0,R)$ for the given network realization whereas $\Ioutc$ corresponds to the transmitters in $\Phid \backslash \mathcal{B}(0,R)$ and $\Phic \backslash \mathcal{B}(0,R)$. Recalling that $\mathbb{E}[K']=K$, the expectation of $\Ioutc$ over the PPPs equals
\begin{align}
\label{eq:I_power3a}
\bIoutc &= \alpha \, \mathbb{E} \left[ \sum_{j=K'+1}^{\infty} \!\!\!\! \Pd \, \rdj^{-\etac} \right] +  \mathbb{E} \left[ \sum_{k=1}^{\infty} \Pc \, \rck^{-\etac} \right] \\
\label{eq:I_power3b}
&= \alpha \,  \int_{R}^{\infty} 2\pi K \lambdab \, \Pd \, r^{1-\etac} {\rm d}r  + \int_{R}^{\infty} 2\pi\lambdab \, \Pc \, r^{1-\etac} {\rm d}r \\
\label{eq:I_power3c}
&= \frac{2\left(\alpha K \Pd + \Pc \right)}{(\etac-2) R^{\etac}}
\end{align}
where (\ref{eq:I_power3b}) follows from Campbell's theorem~\cite[Theorem 4.1]{MHaenggi12} and (\ref{eq:I_power3c}) is obtained by evaluating the integrals and substituting $\lambdab = \frac{1}{\pi R^2}$. 

Similarly, for the D2D link, considering $\mathcal{B}(0,R)$ around the D2D receiver at the origin, the interference power in~(\ref{eq:I_power5b}) can be rewritten as
\begin{align} \label{eq:I_power6}\nonumber
\Id &= \underbrace{\sum_{j=1}^{K'} \Pd \, \rdj^{-\etad} + \alpha \sum_{k=1}^{K''} \Pc \, \rck^{-\etad} }_{\Iind} \\
&\qquad \qquad+  \underbrace{\sum_{j=K'+1}^{\infty} \!\!\!\! \Pd \, \rdj^{-\etad} +  \alpha \!\!\!\!\! \sum_{k=K''+1}^{\infty} \!\!\!\! \Pc \, \rck^{-\etad}}_{\Ioutd}
\end{align}
where $\Iind$ corresponds to the $K'$ D2D transmitters in $\Phid \cap \mathcal{B}(0,R)$ and the $K''$ uplink cellular transmitters in $\Phic \cap \mathcal{B}(0,R)$, whereas $\Ioutd$ corresponds to the transmitters in $\Phid \backslash \mathcal{B}(0,R)$ and $\Phic \backslash \mathcal{B}(0,R)$. Noting that $\mathbb{E}[K']=K$ and $\mathbb{E}[K'']=1$,
the expectation of $\Ioutd$  over the PPPs, computed by applying steps similar to (\ref{eq:I_power3a}) to (\ref{eq:I_power3c}), is 
\begin{align} 
\label{eq:I_power7b}
\bIoutd 
&= \frac{2\left( K\Pd+ \alpha \,\Pc \right)}{(\etad-2) R^{\etad}}.
\end{align}

\subsection{SIR of the Cellular Uplink}
\label{sec: uplink SIR def}

Under the foregoing model for the interference, with power $\Iinc+\bIoutc$, and recalling the intended signal term from (\ref{eq:Signal2c}), the instantaneous SIR of the uplink is
\begin{align}
\label{eq:SIR1}
\SIRc &= \frac{\Pc \, \rco^{-\etac} \, \mathbb{E}\left[|\Hco\bco|^2 \, | \, \Hco \right]}{\Iinc+\bIoutc} \\
\label{eq:SIRc}
&= \rhoc \left| \Hco \right|^2
\end{align}
where the short-term expectation in (\ref{eq:SIR1}) is over $\bco$, conditioned on the fading $\Hco$, while
\begin{align} \label{eq:SIR2}
\rhoc 
&= \frac{\rco^{-\etac}}{\alpha \mu \sum_{j=1}^{K'} \rdj^{-\etac} + \frac{2\left(\alpha \mu K  + 1 \right)}{(\etac-2) R^{\etac}}}
\end{align}
is the local-average SIR at the BS. Further normalizing all the terms by $R^{-\etac}$,
\begin{align}
\rhoc &= \frac{\aco^{-\etac}}{\alpha \mu \sum_{j=1}^{K'} \adj^{-\etac} + \frac{2\left(\alpha \mu K  + 1 \right)}{\etac-2}} \label{messi}
\end{align}
where $\aco = \frac{\rco}{R}$, $\ack = \frac{\rck}{R}$ and $\adj = \frac{\rdj}{R}$ are normalized distances.
Indeed, our formulation is interference limited and therefore invariant to the absolute scale of the network.



\subsubsection{Local-Average SIR Distribution} 

The spatial distribution of the in-circle transmitter locations induces a distribution of its own for $\rhoc$, i.e., a long-term distribution for the local-average SIR, which is presented next separately for underlay and overlay options. 
Though the analytical form of this distribution is unwieldy for underlay ($\alpha=1$) and arbitrary $\etac$, it takes the following closed form for underlay with $\etac = 4$.
\begin{prp}\label{prp_CDFrhoc}
With underlay and $\etac = 4$, the CDF (cumulative distribution function) of $\rhoc$ is

\begin{align}\label{eq:rhocCDF2}
F_{\rhoc}(x) = 
\begin{cases}
\begin{aligned}
&\frac{e^{\kappa^2}\left[\erf\left(\kappa\right)-\erf\left(\frac{\kappa}{\sqrt{1-x}}\right)\right] }{\sqrt{x}}    \\
&\qquad \qquad  + \,  \erf\left(\frac{\kappa\sqrt{x}}{\sqrt{1-x}}\right)
\end{aligned}
	  \quad &  0< x < 1 \\ 
		1 - \frac{e^{\kappa^2} \erfc\left(\kappa\right)}{\sqrt{x}} \quad & x \geq 1
\end{cases}
\end{align}

where $\kappa = \frac{\sqrt{\pi \mu} K}{2}$, $\erf(\nu) = \frac{2}{\sqrt{\pi}} \int_{0}^{\nu} e^{-t^2} {\rm d}t$ is the error function and $\erfc(\nu) = 1-\erf(\nu)$. 
\begin{proof}
	See Appendix~\ref{Appp0}.
\end{proof}
\end{prp}

For overlay ($\alpha =0 $), a form that is both simpler and general in $\etac$ can be obtained.
\begin{prp}\label{prp_CDFrhoc2}
With overlay, the CDF of $\rhoc$ is
\begin{align}
F_{\rhoc}(x) =	1-  \left( \frac{\etac-2}{2 \, x}\right)^{\frac{2}{\etac}} \qquad &  x \geq \frac{\etac-2}{ 2 }	
\end{align}
\begin{proof}
	See Appendix~\ref{Appp00}.
\end{proof}
\end{prp}

\subsubsection{Instantaneous SIR Distribution} \label{sec:inst SIR distri UL}

For a specific network geometry, i.e., given the normalized distances $\aco$ and $\{\adj\}_{j=1}^{K'}$, the value of $\rhoc$ becomes determined. Since $|\Hco|^2$ is exponentially distributed with unit mean, it follows from (\ref{eq:SIRc}) that the cellular uplink $\SIRc$ exhibits a short-term exponential distribution with local-average $\rhoc$ and hence its conditional CDF is
\begin{align}
\label{eq:SIRCDF1}
F_{\SIRc | \rhoc}(\gamma) = 1-e^{-\gamma/\rhoc}.
\end{align}

\subsection{SIR of the D2D Link}
\label{sec: D2D SIR def}

For the D2D link, the instantaneous SIR is
\begin{align}
\label{eq:SIR12}
\SIRd & = \frac{\Pd \, \rdo^{-\etad} \, \mathbb{E}\left[|\Hdo\bdo|^2 \, | \, \Hdo \right]}{\Iind+\bIoutd} \\
\label{eq:SIRd}
&= \rhod \left| \Hdo \right|^2
\end{align} 
where the short-term expectation is over $\bdo$, conditioned on the fading, and
\begin{align} \label{eq:SIR3}
\rhod 
&= \frac{\rdo^{-\etad}}{\sum_{j=1}^{K'} \rdj^{-\etad} + \frac{\alpha}{\mu} \sum_{k=1}^{K''} \rck^{-\etad} + \frac{2\left( K+ \alpha/\mu  \right)}{(\etad-2) R^{\etad}}} \\
&= \frac{\left(\frac{\da}{K^\upbeta}\right)^{-\etad}}{\sum_{j=1}^{K'} \adj^{-\etad} + \frac{\alpha}{\mu} \sum_{k=1}^{K''} \ack^{-\etad} + \frac{2\left( K+ \alpha/\mu  \right)}{\etad-2}} \label{iniesta}
\end{align}
is the local-average SIR at the D2D receiver of interest, 
with $\frac{\da}{K^\upbeta} = \ado = \frac{\rdo}{R} $ and  $\da = \frac{\dr}{R}$.


\subsubsection{Local-Average SIR Distribution}

\begin{prp}\label{prp_CDFrhod}
The CDF of $\rhod$ is
\begin{align}\label{eq:rhodCDF}\nonumber
F_{\rhod}(x) 
&= \frac{1}{\pi} \sum_{k=1}^{\infty} \left[ \frac{x^{2/\etad} \, \da^2}{K^{2\upbeta}}  \, \left(K+ \frac{\alpha}{\mu^{2/\etad}}\right) \, \Gamma\left(1-\frac{2}{\etad}\right)\right]^k \\ 
&\qquad \qquad \quad  \cdot \frac{\Gamma\left(\frac{2k}{\etad}\right)}{k!} \, \sin\left[k \pi \left(1-\frac{2}{\etad}\right) \right]
\end{align}
where $\Gamma(\cdot)$ is the Gamma function. For $\etad = 4$, the above reduces to
\begin{align}\label{eq:rhodCDF2}
F_{\rhod}(x) 
= \erf\left[\frac{\sqrt{\pi \, x} \, \da^2}{2 \, K^{2\upbeta}}  \, \left(K+ \frac{\alpha}{\sqrt{\mu}}\right) \right].
\end{align}
\begin{proof}
	See Appendix~\ref{App5}.
\end{proof}
\end{prp}


\subsubsection{Instantaneous SIR Distribution}

Given $\ado = \frac{\da}{K^\upbeta}$, $\{\adj\}_{j=1}^{K'}$ and $\{\ack\}_{k=1}^{K''}$, the value of $\rhod$ becomes determined and it follows from (\ref{eq:SIRd}) that
\begin{align}
\label{eq:SIRCDF3}
F_{\SIRd | \rhod }(\gamma) = 1-e^{-\gamma/\rhod}.
\end{align}

\section{Link Spectral Efficiency}
\label{sec:Seff}

We now turn our attention to the ergodic spectral efficiency, arguably the most operationally relevant quantity in contemporary systems~\cite{Loz-Jin12}. 


\subsection{Specific Network Geometry}
\label{sec:Seff for abs geo}

For given $\rhoc$, the spectral efficiency of the corresponding cellular uplink is
\begin{align}
\SEc(\rhoc) &= \mathbb{E}\left[ \log_2(1+\SIRc | \rhoc)\right]  \\ 
&= \int_{0}^{\infty}\log_2(1+\gamma) \, {\rm d}F_{\SIRc | \rhoc}(\gamma) \\
\label{eq:SEC}
&= e^{1/\rhoc}E_1\left(\frac{1}{\rhoc}\right)\log_2e
\end{align}
where $E_1(\zeta) = \int_{1}^{\infty}t^{-1}e^{-\zeta t}{\rm d}t$ is an exponential integral.

Similarly, for given $\rhod$, the spectral efficiency of the corresponding D2D link equals
\begin{align}
\SEd(\rhod) 
\label{eq:D2DSE1}
&= e^{1/\rhod}E_1\left(\frac{1}{\rhod}\right)\log_2e.
\end{align}


\begin{exmp}
\label{example2}

\begin{figure}[h]  
\centering
 \includegraphics [width=0.97\columnwidth] {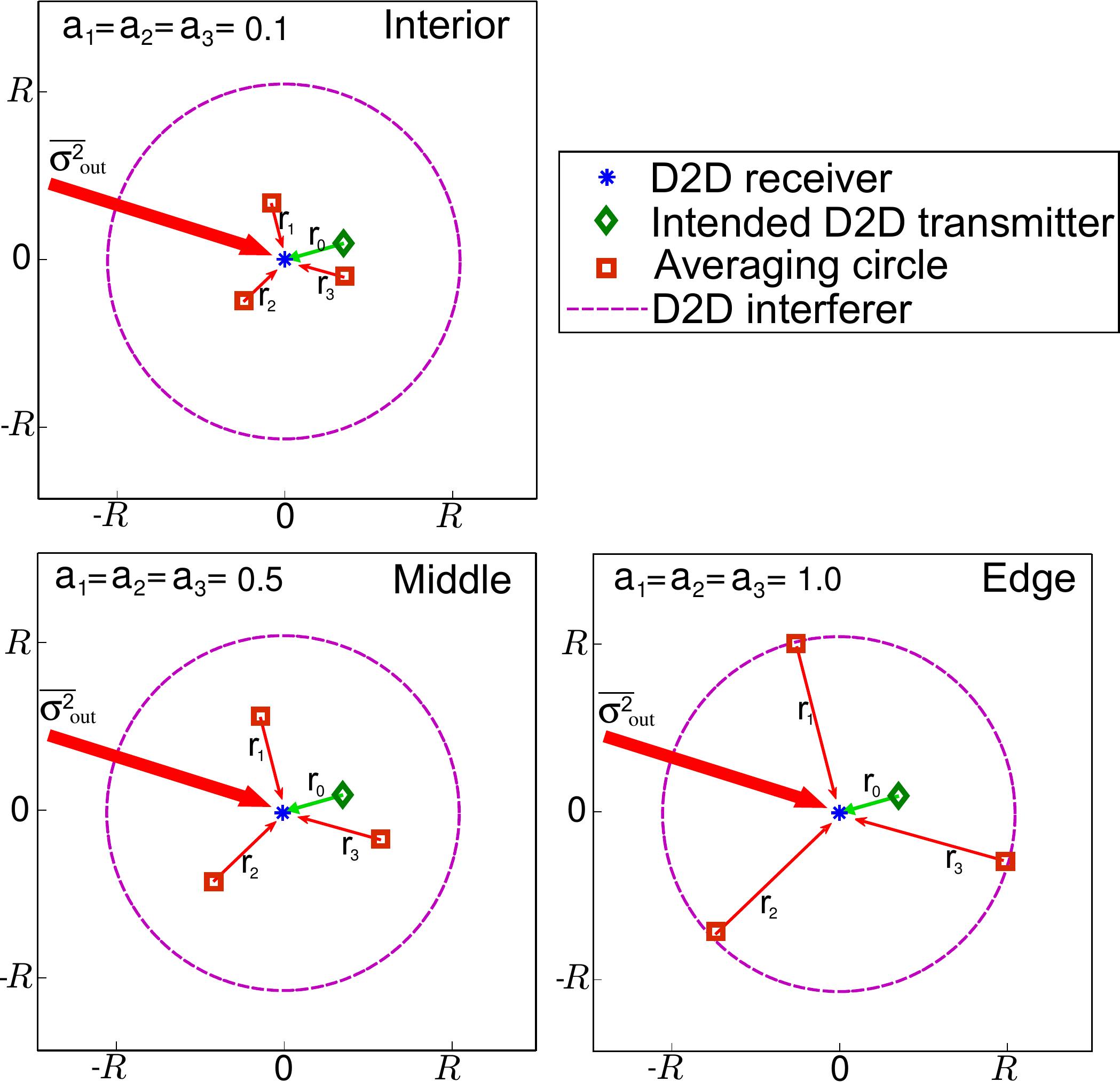}
\caption{The three situations considered in Example \ref{example2}.}
\label{Fig:val4}
\end{figure}

Consider a D2D link in an overlay system with $K=3$ links per cell on average, with $\upbeta=0$ (i.e., with link lengths that are independent of the user density)
and with the pathloss exponent $\etad = 4.5$.
Draw around the receiver an averaging circle of size $R$ such that only the locations of the inner interferers are conditioned on.
Compared in Fig.~\ref{Fig:val3} are the link spectral efficiency in (\ref{eq:D2DSE1}) and the exact mutual information under the non-Gaussian interference $\zdo$ as per (\ref{eq:zintd}), with such mutual information numerically computed
through lengthy Monte-Carlo histograms
and averaged over many fading realizations for many snapshots of the interferer locations outside the averaging circle. Three different situations are considered, with the interferers placed at the interior, middle and edge of the averaging circle (cf. Fig. \ref{Fig:val4}).
Excellent matches are observed, supporting our interference modeling approach.

\begin{figure}[h]  
\centering
 \includegraphics [width=0.97\columnwidth]{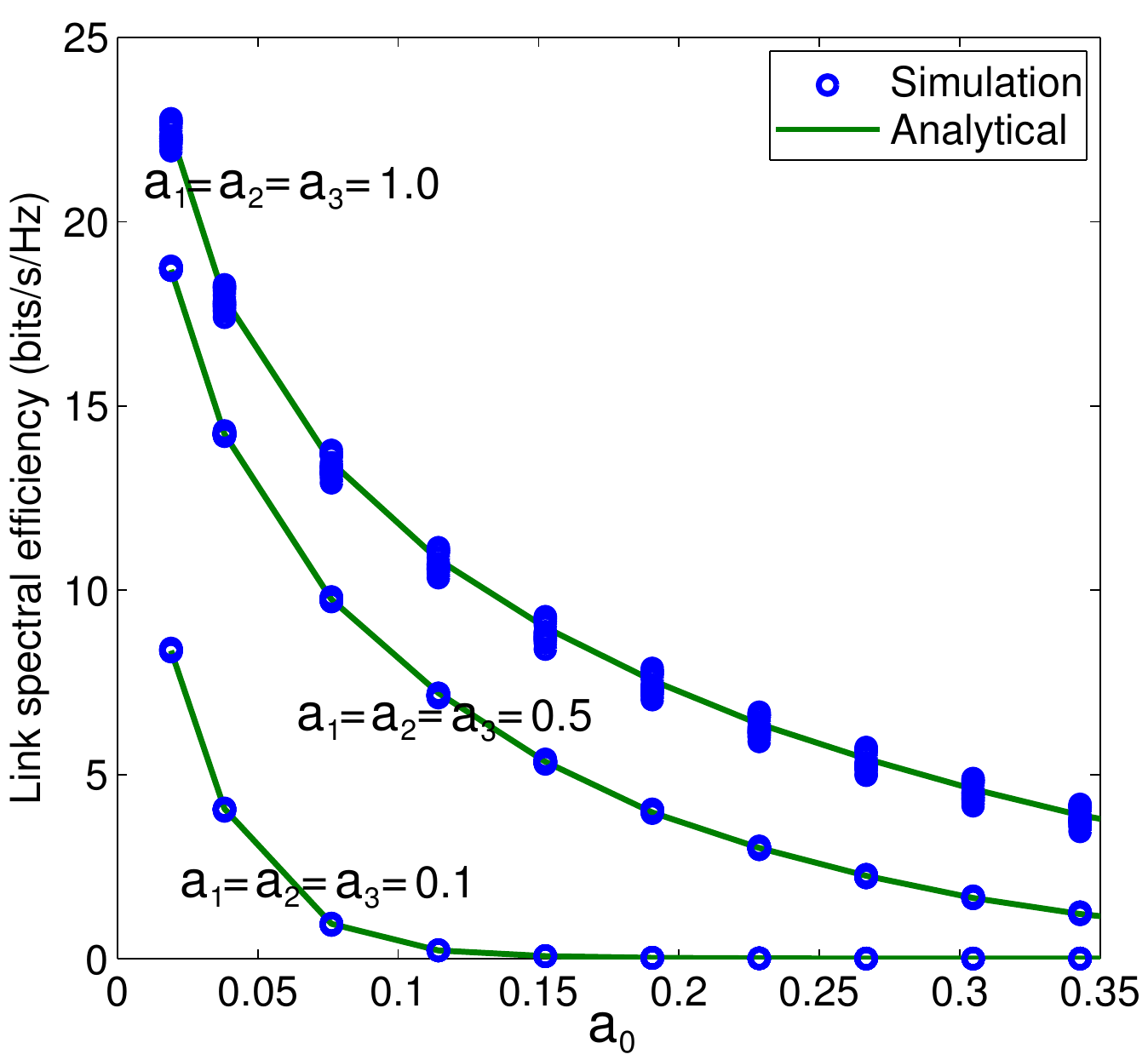}
\caption{D2D link spectral efficiency in an overlay system as function of the normalized link distance $\ado$ with $\upbeta = 0$ and $K = 3$: analysis with our interference model vs. simulation with the interference modeled as per  (\ref{eq:zintd}). 
The pathloss exponent $\etad = 4.5$.}
\label{Fig:val3}
\end{figure}
\end{exmp}

Similarly good agreements are observed for other overlay settings and also for underlay settings.
In fact, as mentioned earlier, the spectral efficiency obtained with our analytical approach is not only an accurate approximation to the value obtained with the interference modeled as per (\ref{eq:zintd}), but arguably more operationally relevant because it is unlikely that the receiver can learn the exact distribution of such interference and, even if it could, a standard decoder for Gaussian noise might be featured.


Moving beyond the specific placements in Example \ref{example2}, the spatial distribution of the transmitters induces, through $\rhoc$ and $\rhod$, a distribution of their own for $\SEc$ and $\SEd$. 
Provided $F_{\rhoc}(\cdot)$ and $F_{\rhod}(\cdot)$, the distributions of $\SEc$ and $\SEd$ can be evaluated numerically, using (\ref{eq:SEC}) and (\ref{eq:D2DSE1}), for given values of the parameters.
However, characterizing them analytically appears challenging because of the exponential integral function in the spectral efficiency expressions.
Alternatively, invoking \cite{Catreux02}
\begin{align}
 e^{x} \, E_1(x) \, \log_2 e \approx 1.4 \log_e \left(1+\frac{0.82}{x}\right)
\end{align}
we can approximate the CDFs of $\SEc$ and $\SEd$ as
\begin{align}\label{eq:SECDF_approx2}
F_{\SEc}(\nu) &\approx F_{\rhoc}\left(\frac{e^{\frac{\nu}{1.4}}-1}{0.82}\right) \\
\label{eq:SECDF_approx}
F_{\SEd}(\nu) &\approx F_{\rhod}\left(\frac{e^{\frac{\nu}{1.4}}-1}{0.82}\right).
\end{align}

\begin{exmp}\label{SECDF_valid}
In Fig.~\ref{Fig:val5}, the approximated CDFs in (\ref{eq:SECDF_approx2}) and (\ref{eq:SECDF_approx}) are contrasted against the ones obtained numerically by means of the corresponding local-average SIR CDFs (Props. \ref{prp_CDFrhoc}, \ref{prp_CDFrhoc2} and \ref{prp_CDFrhod}) and the link spectral efficiency equations, and the ones obtained completely through Monte-Carlo. Setting $K = 10$, $\da = 0.1$, $\etac = 4$ and $\etad = 4.5$, underlay with $\upbeta = 0.25$ and overlay with $\upbeta = 0.5$ are considered. Very good agreements are observed, again validating our interference modeling approach.
\begin{figure}[h]  
\centering
 \includegraphics [width=0.97\columnwidth]{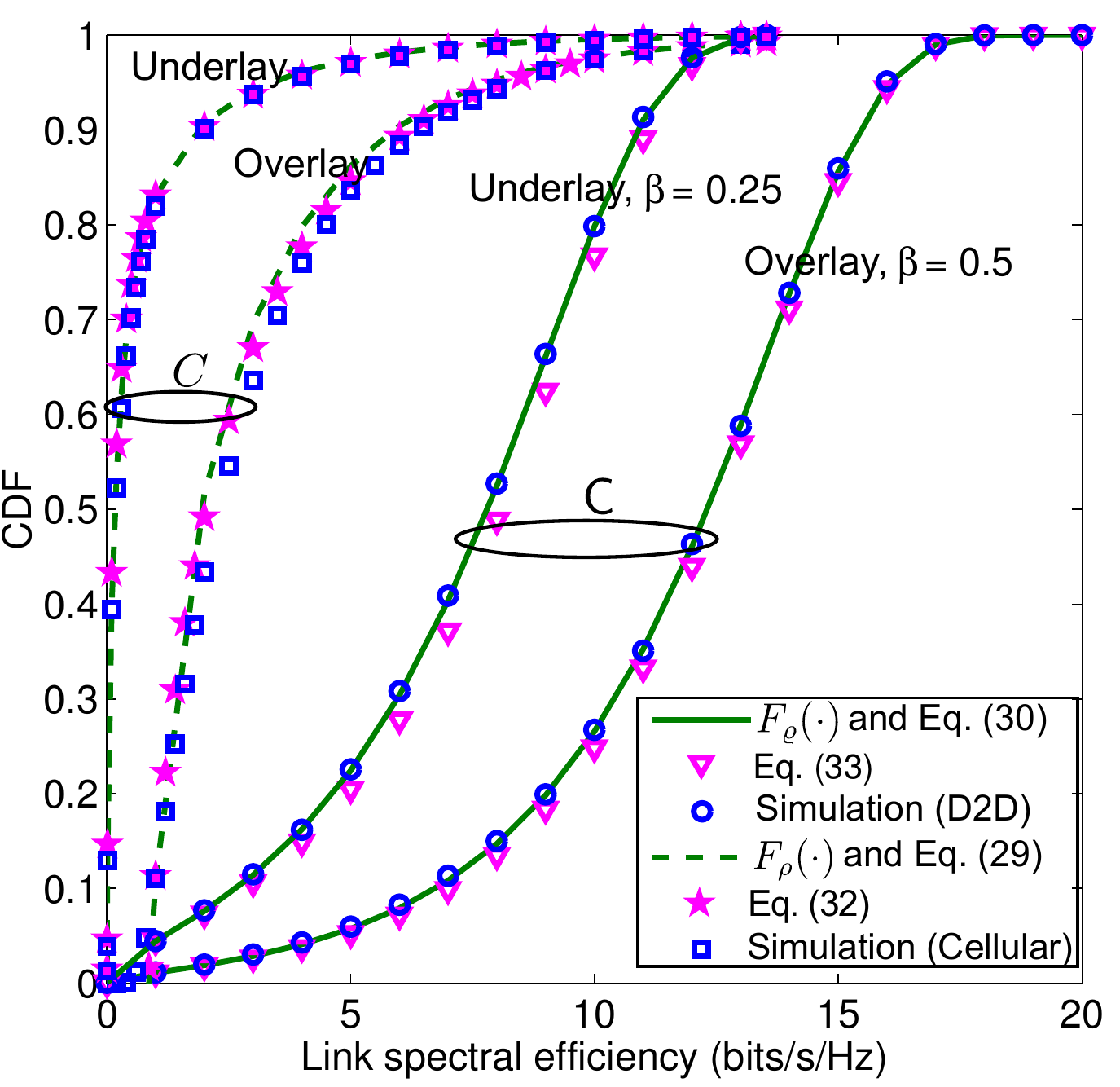}
\caption{CDFs of $\SEc$ and $\SEd$ for $K = 10$, $\da = 0.1$, $\etac = 4$ and $\etad = 4.5$.}
\label{Fig:val5}
\end{figure}
\end{exmp}

\subsection{Average Network Geometry}\label{sec:Average_Net}
\label{sec:Seff for avg nw geo}


As an alternative to the characterization for specific network geometries presented in the previous section, we can choose to characterize the average spectral efficiency over all possible such geometries. Although, as mentioned earlier, the quantities thus obtained---favorite outcomes in stochastic geometry---are less informative, they do allow calibrating system-level benefits.

To effect the spatial averaging, slightly different approaches are computationally more convenient in our framework depending on the type of link (cellular/D2D) and on the case (underlay/overlay). Hence, each is separately presented next, with the details relegated to the Appendix.


\subsubsection{Cellular Uplink}
\label{sec:uplink Seff for avg nw geo}

\begin{prp}
\label{prp_avg_cellSE3} 
With underlay, the cellular uplink spectral efficiency averaged over all network geometries is
 \begin{align}
 \label{eq:conditional_avSE}
  \bSEc =  2  \int_{0}^{\infty} \frac{\log_2 e}{\gamma+1} \,  \int_{0}^{1} a \, e^{-\gamma \frac{2\,a^{\etac}}{\etac-2}\,-\,(\gamma\mu)^{\frac{2}{\etac}} a^2 \, K \, \Gamma\left(1-\frac{2}{\etac}\right)} {\rm d}a \,  {\rm d}\gamma
 \end{align}
 which, for $\etac = 4$, simplifies to
  \begin{align}
  \label{eq:underlay_avcellSE}\nonumber
 \bSEc &= \frac{\sqrt{\pi}e^{-\pi\mu K^2/4}}{2 \log_e 2} \\
 &\quad \cdot \int_{0}^{\infty} \frac{  \erf\left(\sqrt{\gamma}+\frac{\sqrt{\pi\mu}K}{2}\right) - \erf\left(\frac{\sqrt{\pi\mu}K}{2}\right) }{\sqrt{\gamma} \, (1+\gamma)}  \, {\rm d}\gamma
\end{align}
\begin{proof}
See Appendix~\ref{App4}.
\end{proof}
\end{prp} 

With overlay ($\alpha=0$),
a compact result involving only the Meijer-G function 
\begin{align}
G^{m,n}_{p,q}\left(z \Biggl | \Biggr. \begin{array}{l l} a_1,...,a_n,a_{n+1},...,a_p \\ b_1,...,b_m,b_{m+1},...,b_q \end{array} \right)
\end{align}
and the cellular pathloss exponent $\etac$ is obtained.

\begin{prp}
\label{prp_avg_cellSE1}
With overlay, the cellular uplink average spectral efficiency over all network geometries is
\begin{align} \label{eq:UL seff Overlay}
\bSEc = \frac{2\log_2e}{\etac} \, G^{2,2}_{2,3} \left(\frac{2}{\etac-2} \Biggl | \Biggr. \begin{array}{l l} 0,\frac{\etac-2}{\etac} \\ 0,0,\frac{-2}{\etac} \end{array} \right).
\end{align}
\begin{proof}
See Appendix~\ref{App3}.
\end{proof}
\end{prp}

The versatility of our interference modeling approach is in full display here, facilitating a closed-form expression for a quantity that had previously been obtained only in integral form.


\subsubsection{D2D link}
\label{sec:D2D Seff for avg nw geo}



\begin{prp}
\label{avg_D2DSE1}
The D2D link spectral efficiency averaged over all network geometries equals
\begin{align}\label{eq:underD2Dav}
\bSEd &=  \int_{0}^{\infty} \frac{\log_2 e}{\gamma+1} \, e^{-\gamma^{2/\etad} \, \frac{\da^2}{K^{2\upbeta}}  \, \left(K+ \frac{\alpha}{\mu^{2/\etad}}\right) \, \Gamma\left(1-\frac{2}{\etad}\right)} \, {\rm d}\gamma
\end{align}
which, for $\etad = 4$, reduces to
\begin{align}
\bSEd &= 2  \left[ \sin \left( \mathcal{K} \, \da^2 \right) {\rm si} \left( \mathcal{K} \, \da^2 \right) -  \cos \left( \mathcal{K} \, \da^2 \right){\rm ci} \left( \mathcal{K} \, \da^2 \right) \right] \log_2 e
 \label{eq:Seff D2D eta4}
\end{align}
where $\mathcal{K}=\frac{\sqrt{\pi}}{K^{2\upbeta}}(K+\frac{\alpha}{\sqrt{\mu}})$ while the trigonometric integrals ${\rm si}(\cdot)$ and ${\rm ci}(\cdot)$ are respectively given by ${\rm si}(x) = \int_{x}^{\infty}\frac{\sin(t)}{t} {\rm d}t $ and ${\rm ci}(x) = -\int_{x}^{\infty}\frac{\cos(t)}{t} {\rm d}t$.
\begin{proof}
	See Appendix~\ref{App1}.
\end{proof}
\end{prp}

A particularly interesting special case arises when $\upbeta = 1/2$, as then the D2D link length is $\ado \propto \frac{1}{\sqrt{K}}$ and the interferer link distances $\{\adj\}_{j=1}^{\infty}$ to the points in the PPP $\Phid$ satisfy $\mathbb{E}[\adj] \propto \frac{1}{\sqrt{K}}$~\cite{Haenggi05}. Intuitively, one would expect the dependence on the user density to vanish, and that is indeed the case with overlay.

\begin{cor}
\label{prp_avg_D2DSE2_ovr}
With overlay and $\upbeta=1/2$, the D2D link spectral efficiency averaged over all network geometries is
 \begin{align} \label{eq:D2DavgSE}
 \bSEd =  \int_{0}^{\infty} \frac{ \log_2 e}{\gamma+1} \, e^{-\gamma^{2/\etad} \, \da^2 \, \Gamma\left(1-\frac{2}{\etad}\right)} \, {\rm d}\gamma
 \end{align}
 which, for $\etad=4$, reduces to
 \begin{align}
 \bSEd &= 2  \left[ \sin \left(\sqrt{\pi}\,\da^2 \right) {\rm si} \left( \sqrt{\pi}\,\da^2 \right) \right. \nonumber \\
 &\qquad \qquad \left. - \cos \left( \sqrt{\pi} \, \da^2 \right){\rm ci} \left( \sqrt{\pi} \, \da^2 \right) \right] \log_2 e.
 \end{align}
\end{cor}

\begin{exmp}

\begin{figure}[h]  
\centering
\includegraphics [width=0.97\columnwidth]{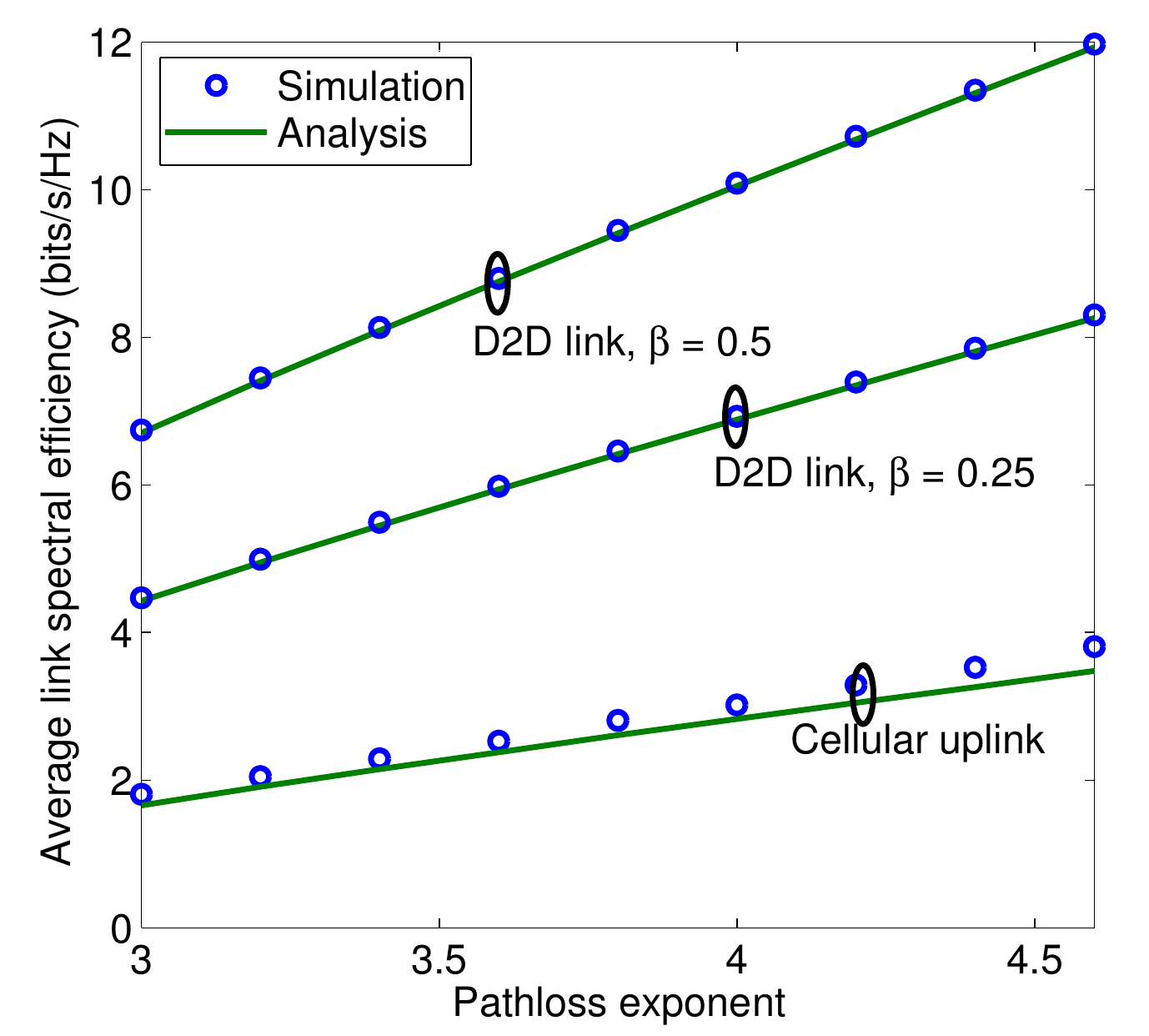}
\caption{Average spectral efficiencies of uplink and D2D link for varying $\etac$ and $\etad$ with overlay.}
  \label{Fig:Exm2}
\end{figure}

For an overlay system ($\alpha=0$) with $\da = 0.1$ and $K=10$, Fig.~\ref{Fig:Exm2} shows the average spectral efficiencies in~(\ref{eq:UL seff Overlay}) and~(\ref{eq:underD2Dav}) alongside the respective numerically computed values with $\zco$ as in~(\ref{eq:zintc}) and $\zdo$ as in~(\ref{eq:zintd}).
The match is excellent, once again evincing the goodness of our interference modeling approach. 
\end{exmp}
 

\section{Benefits of D2D}
\label{sec:Benifits of D2D-cellular}

In order to gauge the benefits of D2D and to demonstrate the usefulness of the framework developed in earlier sections, we next provide some examples. 
Unless otherwise specified, the pathloss exponents are $\etac=3.5$ and $\etad = 4.5$.


\subsection{Overlay D2D}
\begin{figure}  
\centering
\includegraphics [width=0.99\columnwidth]{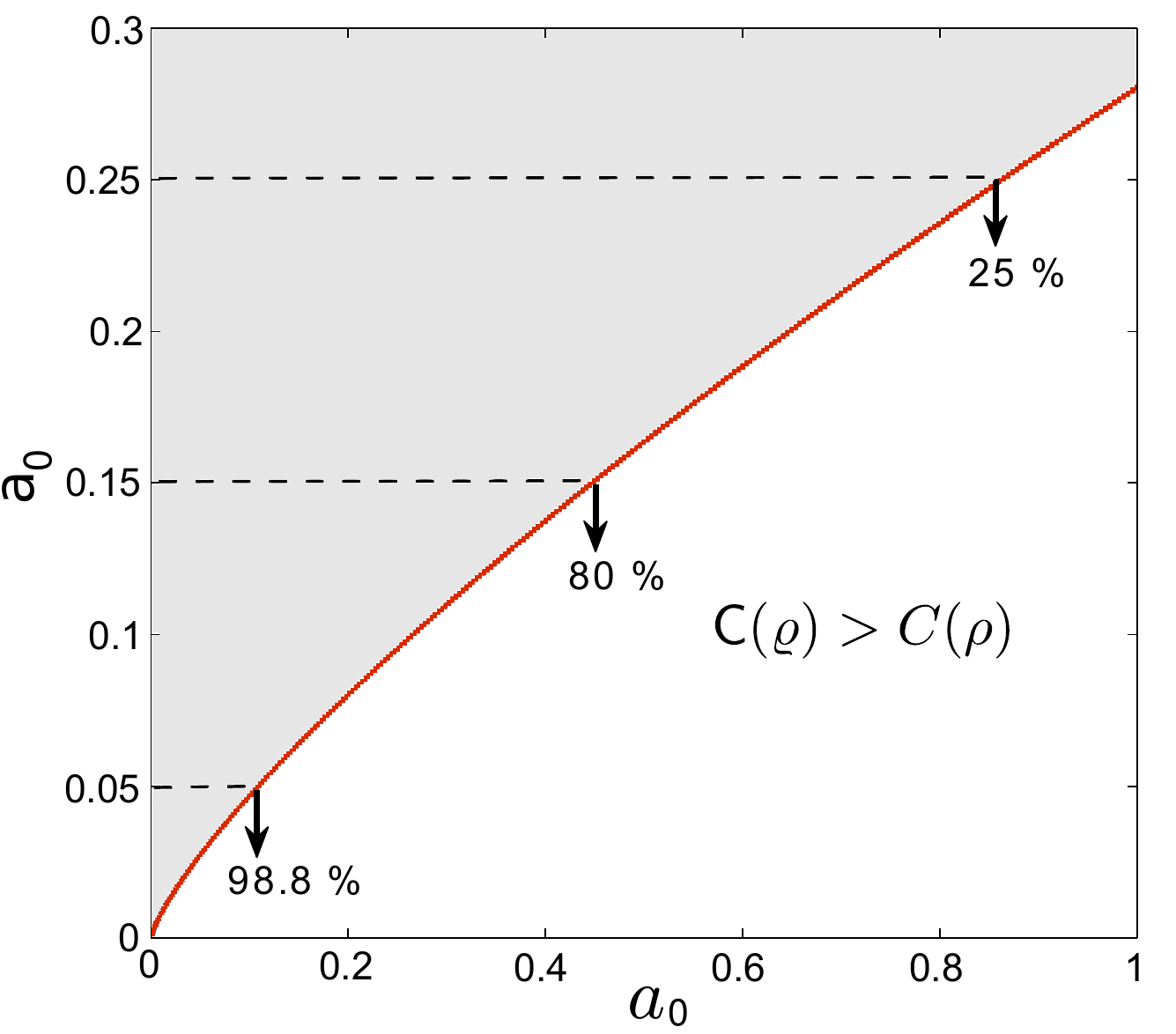}
 \caption{Contour plot for the relationship in (\ref{barcelona}). Within the unshaded region, $\SEd(\rhod)>\SEc(\rhoc)$.
 }
  \label{Fig:Exm3}
\end{figure}

\begin{exmp}
\label{messi73}
Consider an overlay system with fixed D2D link distance $\ado = \da$ ( i.e., with $\upbeta = 0$) and
with an average of $K=10$ D2D links per cell.
To render the system as typical as possible, $K'$ is set to its expected value and, for $j=1,\ldots,K'$, $\adj$ is set to the expected value of the normalized distance to the $j$th nearest neighboring point in a PPP with density $\lambdad$~\cite{Haenggi05}. 
In such a setup, equating the link spectral efficiencies $\SEc(\rhoc)$ and $\SEd(\rhod)$ we obtain 
\begin{align}
\frac{(3.5-2) \, \aco^{-3.5}}{2} &= \frac{\ado^{-4.5}}{\sum_{j=1}^{10} \left(\frac{\Gamma(0.5+j)}{\sqrt{10} \, \Gamma(j)}\right)^{-4.5} + \frac{2 \cdot 10}{4.5-2}} 
\end{align}
which simplifies into
\begin{align}
\label{barcelona}
\aco &= 0.512 \, \ado^{\frac{4.5}{3.5}}
\end{align}
for which a contour plot is shown in Fig.~\ref{Fig:Exm3}.
Within the unshaded region, the D2D link has a better spectral efficiency than a corresponding uplink transmission from the same user would have, and thus D2D is advantageous.
The share of geometries for which $\SEd(\rhod)>\SEc(\rhoc)$ for a given $\ado$ is  $\mathbb{P}[\aco>x] = 1-x^2$ where $x$ is the corresponding $x$-axis value of the contour. Some such shares are displayed, e.g., for $\ado=0.15$ D2D is preferable in $80\%$ of situations.
\end{exmp} 

Example \ref{messi73} shows how, from a link vantage, D2D is very often preferable to communicating via the BS \emph{even if only the uplink is considered}. With the resource costs of both uplink and downlink considered, the appeal of D2D would increase even further.

\begin{exmp}
\begin{figure}  
\centering
 \includegraphics [width=0.97\columnwidth]{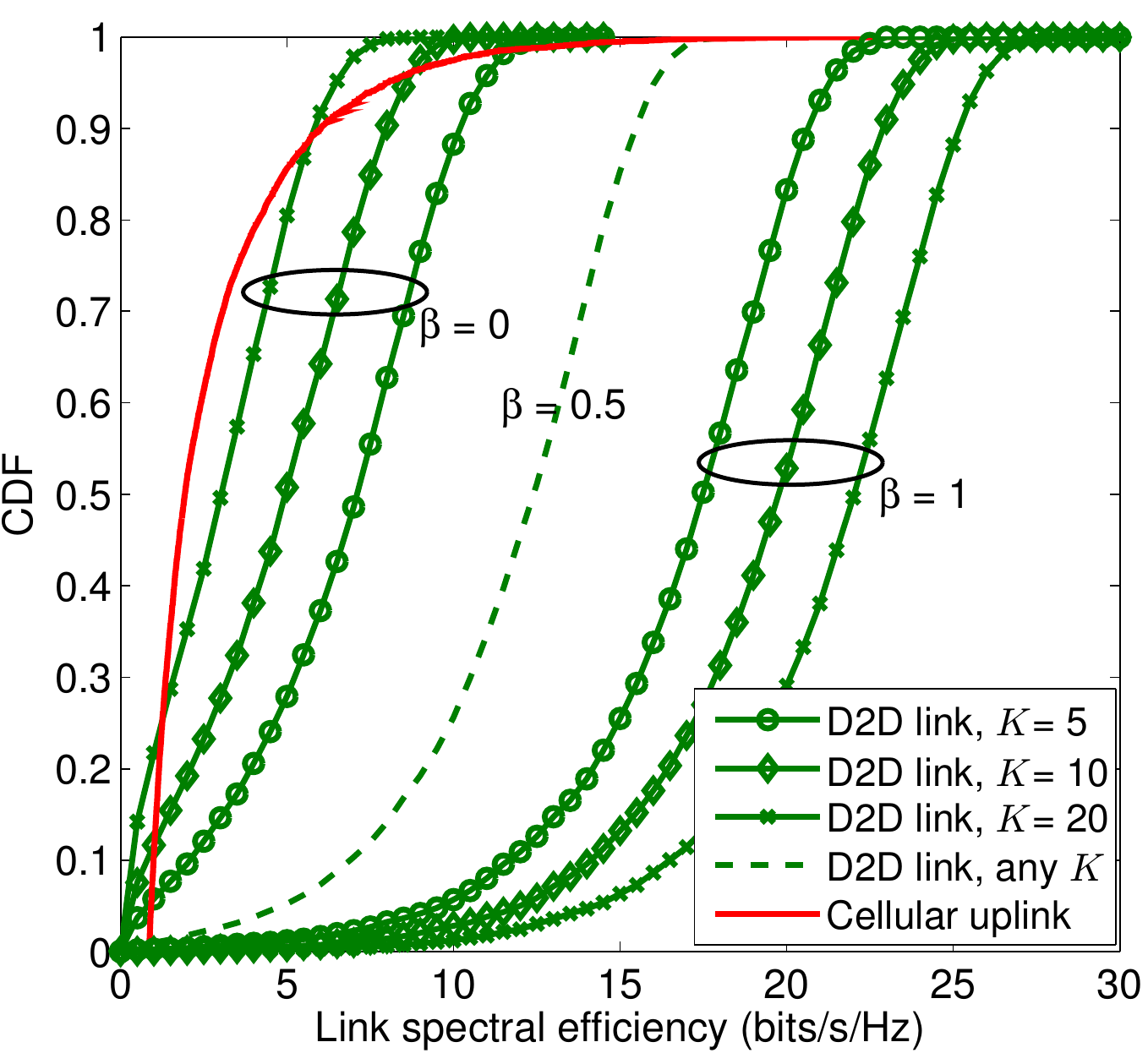}
 \caption{CDFs of cellular and D2D link spectral efficiencies in an overlay system with $\da = 0.1$ and different $\upbeta$.}
 \label{Fig:Exm4}
\end{figure}

Considering overlay and $\da = 0.1$, CDFs of $\SEc(\rhoc)$ and $\SEd(\rhod)$ are plotted in Fig.~\ref{Fig:Exm4} for various $\upbeta$ and $K$. Even for high densities $K$, with the system brimming with D2D interference, thanks to their short range many D2D links enjoy higher spectral efficiencies than the corresponding cellular uplink. We observe the following from the CDFs of $\SEd(\rhod)$:
\begin{itemize}
\item When $\upbeta < 1/2$, $\SEd(\rhod)$ worsens with increasing $K$.
\item When $\upbeta = 1/2$, $\SEd(\rhod)$ is independent of $K$.
\item When $\upbeta > 1/2$, $\SEd(\rhod)$ improves with increasing $K$.
\end{itemize}
The above conforms with intuition as the interferer distances $\{\adj\}_{j=1}^{\infty}$ shrink with $K$ on average, $\mathbb{E}[\adj] \propto \frac{1}{\sqrt{K}}$. When $\upbeta < 1/2$, $\SEd(\rhod)$ decreases with $K$ as $\ado$ shrinks slower than the interferer distances while, when $\upbeta > 1/2$, $\SEd(\rhod)$ increases with $K$ as $\ado$ shrinks faster than the interferer distances. 
\end{exmp} 

The number of D2D links that can coexist on a given signaling resource is large, and to better appreciate the benefits of such dense spectral reuse we next turn our attention to the system spectral efficiency (bits/s/Hz per cell), which reflects the benefits of this reuse.

\begin{exmp}
\label{champions}

\begin{figure}  
\centering

 \includegraphics [width=0.975\columnwidth]{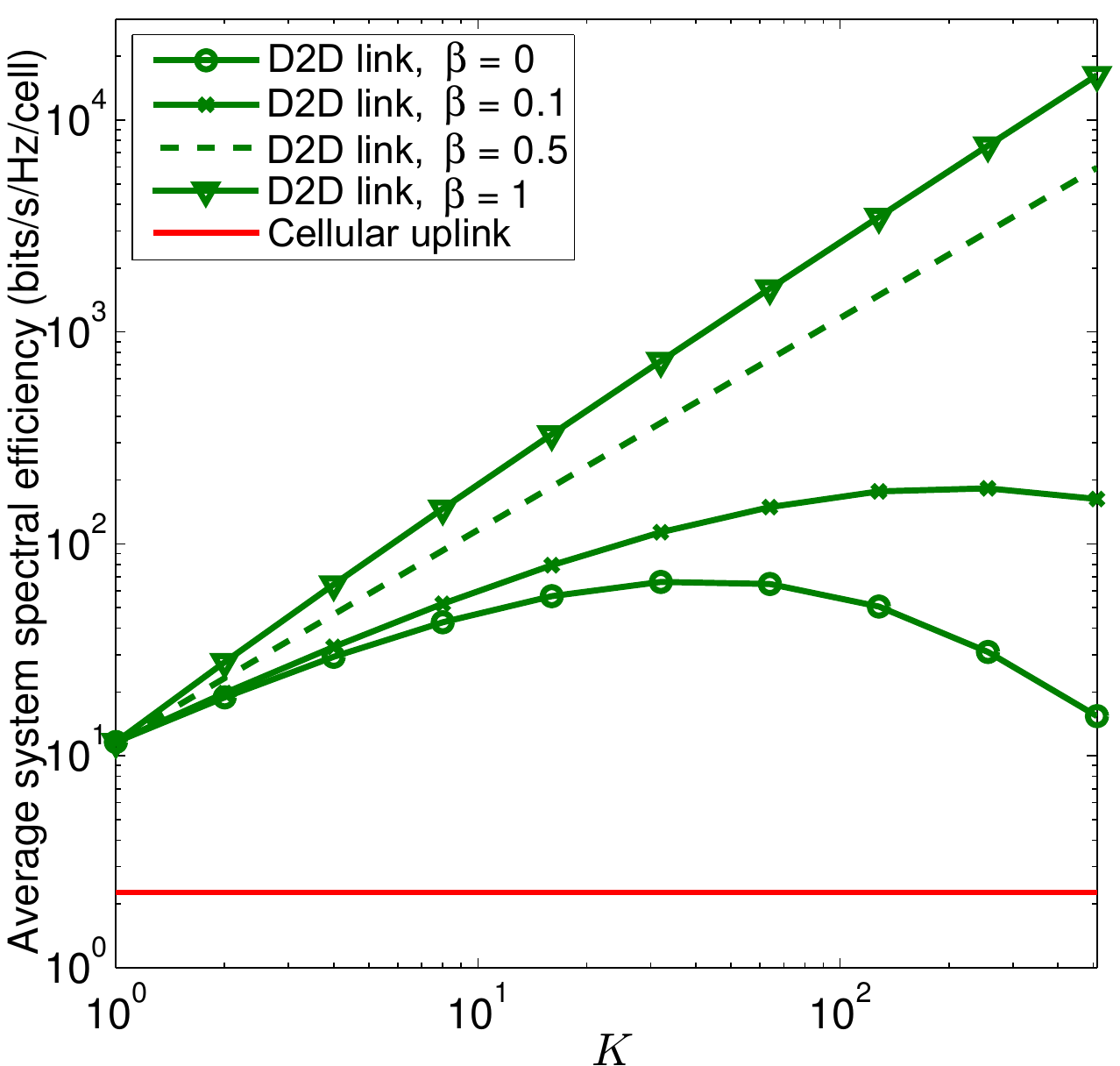}
 \caption{Average system spectral efficiency in an overlay system with $\da=0.1$.}
  \label{Fig:Exm6}
\end{figure}

Since there are $K$ active D2D links per cell on average, the average system spectral efficiency of the D2D traffic is $K\bSEd$ whereas, for the cellular uplink, the average system spectral efficiency is $\bSEc$ as there is only one active cellular user per cell. Shown in Fig.~\ref{Fig:Exm6} is the comparison of these quantities as function of $K$, for various $\upbeta$.
As $K$ grows beyond the range of values shown in the plot, the D2D link curves for $\upbeta = 0$ and $\upbeta = 0.1$ eventually fall below the cellular uplink curve.
The following is observed for a variety of such settings:
\begin{itemize}
\item For each $\upbeta < 1/2$, there is an optimum ``load'' $K$. 
\item When $\upbeta \geq 1/2$, the D2D system spectral efficiency increases monotonically with $K$.
\item Even when not monotonic in $K$, the D2D system spectral efficiency is generally \emph{much} higher than its cellular counterpart.
\end{itemize}
\end{exmp}



\subsection{Underlay D2D}


%
%

\begin{exmp}


\label{example_underlay2}
\begin{figure}[h]  
\centering
\includegraphics [width=0.98\columnwidth]{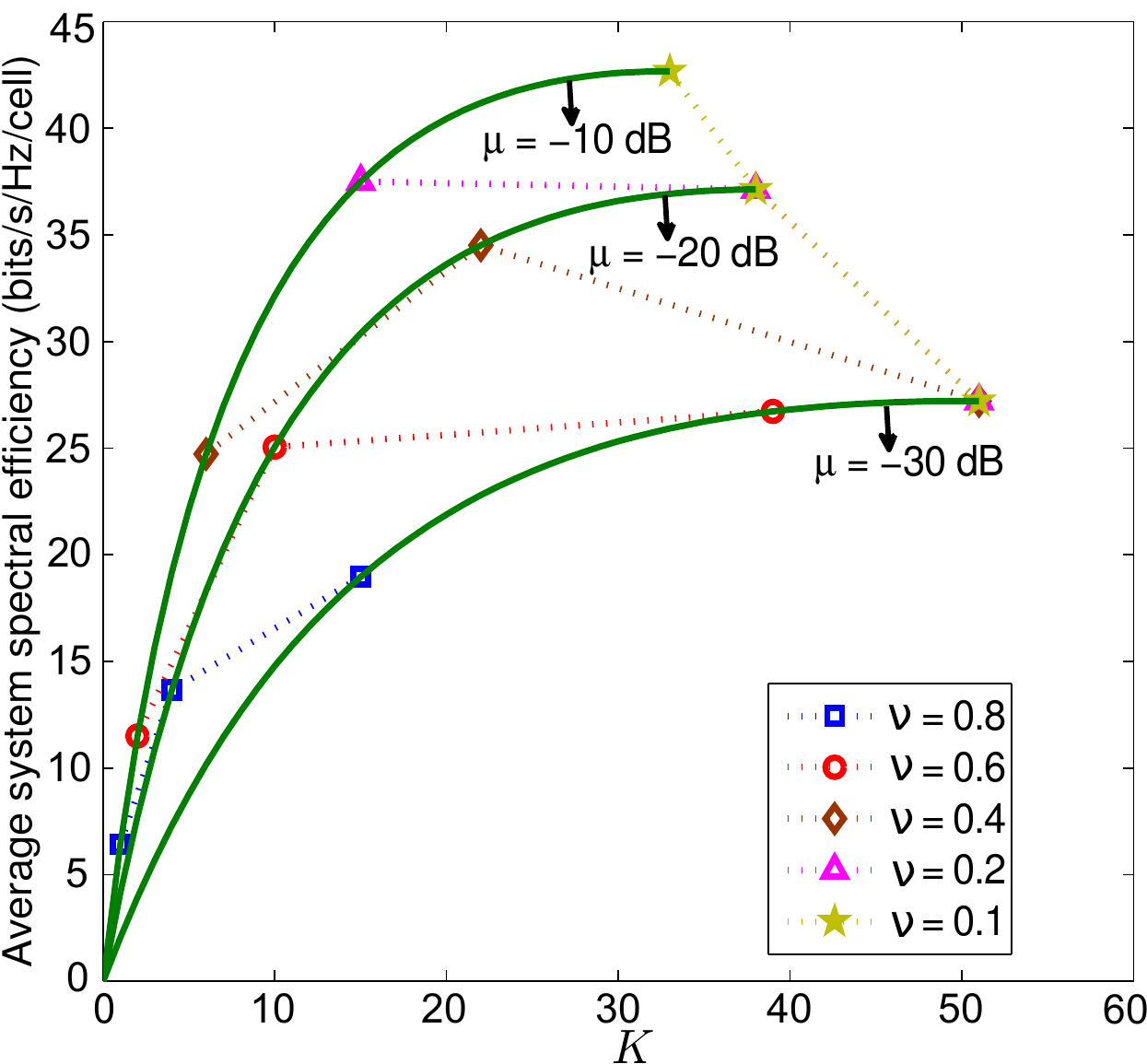}
 \caption{Average system spectral efficiency of underlaid D2D with $\da = 0.12$ and $\upbeta=0$ for different values of $K$, $\mu$ and $\upnu$.}
 \label{Fig:Exm56}
\end{figure}

In Fig. \ref{Fig:Exm56}, the average system spectral efficiency (bits/s/Hz per cell) achieved by underlaid D2D with $\ado = 0.12$ is plotted until its peak value by varying $K$, for different values of $\mu$. 
Suppose that we want the average uplink system spectral efficiency to satisfy $\bSEc \geq \upnu \, \bSEc|_{K=0}$ where $\bSEc|_{K=0}$ denotes its value without D2D and $\upnu>0$ parametrizes its degradation; for instance, $\upnu = 0.8$ means less than 20\% degradation. The maximum average system spectral efficiencies of D2D for different values of $\upnu$ are indicated in the figure. 
\end{exmp}

The strong dependence on $\upnu$ in Example \ref{example_underlay2} indicates that, with underlay, the cellular uplink spectral efficiency is severely affected by D2D interference;
this encourages us to look into ways of protecting the cellular uplink from the D2D interferers, which is the focus of the next section.

\section{Underlaid D2D with Exclusion Regions}

One way to reduce the interference seen in the uplink with underlay is to have exclusion regions around the BSs wherein no co-channel D2D transmitters are allowed, an idea explored for related settings in~\cite{SyuLeeD2D13,MNiD2D15}.
  
\begin{figure}[h]  
\centering
\includegraphics [width=0.97\columnwidth]{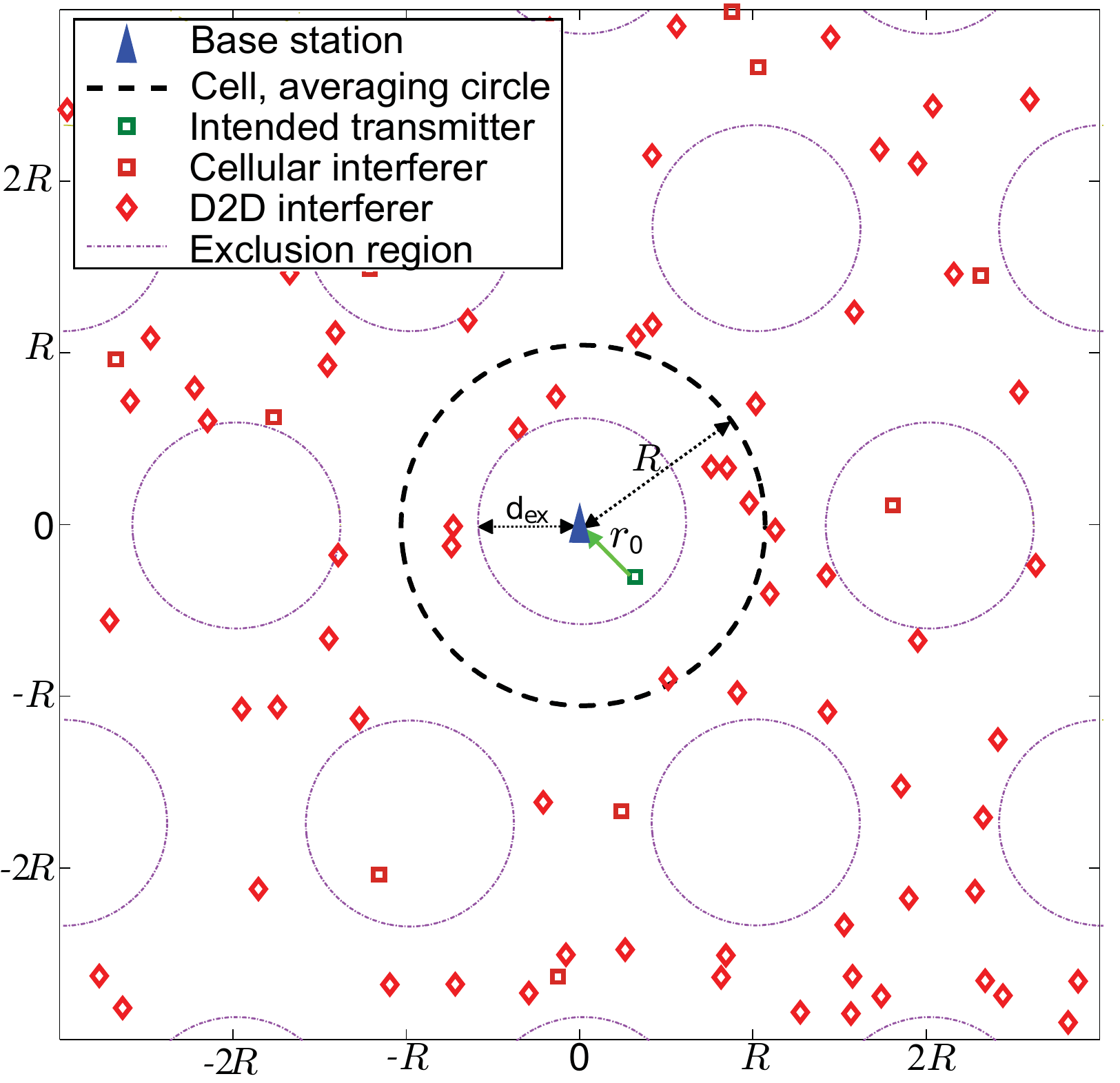}
 \caption{Cellular uplink with no underlaid D2D transmitters allowed in circular exclusion regions of radius $\dc$ around the BSs. At the origin is a receiving BS and shown with a square marker within the circle of radius $R$ is its intended cellular transmitter; shown with square markers outside the circle are the cellular interferers; shown with diamond markers are the D2D interferers.}
  \label{Fig:exclngrid}
\end{figure}

Let us consider circular exclusion regions of radius $\dc$ (cf. Fig.~\ref{Fig:exclngrid}) and normalized radius $\adc = \frac{\dc}{R}$.
With the introduction of such exclusion regions, the D2D interferer locations no longer conform to a homogeneous PPP, which makes the analysis cumbersome. To circumvent this difficulty, we model the D2D interferer locations outside the averaging circle $\mathcal{B}(0,R)$ as belonging to a different homogeneous PPP $\Phidt$ with a scaled-down density $\lambdadt = p \lambdad$, where $p = 1-\adc^2$ such that $\lambdadt$ coincides with the average number of active D2D transmitters per unit area;
the goodness of this model is validated in a later example.
The interference power emanating from outside $\mathcal{B}(0,R)$ is averaged over the locations of those interfererers, which for the cellular uplink gives 
\begin{align}
\bIoutc &=\frac{2\left(p K \Pd + \Pc \right)}{(\etac-2) R^{\etac}}
\end{align}
and for the D2D link gives
\begin{align}
\bIoutd &= \frac{2\left(p K \Pd+ \Pc \right)}{(\etad-2)R^{\etad}}.
\end{align}

For the cellular uplink, the interference power from the transmitters inside $\mathcal{B}(0,R)$ is 
\begin{align}
\Iinc &=  \sum_{j=1}^{\mathcal{K}'} \Pd \, \rdj^{-\etac}
\end{align}
where the $\mathcal{K}'$ D2D transmitters are located within an annulus with inner radius $\dc$ and outer radius $R$ denoted by $\mathcal{A}(\dc,R)$. The locations of the D2D transmitters within $\mathcal{A}(\dc,R)$ conform to the points of the PPP $\Phid$ with density $\lambdad$.

As of the D2D link, recall that to study it we shift the origin to the D2D receiver under consideration.
The interference power from the transmitters inside $\mathcal{B}(0,R)$ is
\begin{align}
\Iind &=  \sum_{j=1}^{\mathcal{K}''} \Pd \, \rdj^{-\etad} + \sum_{k=1}^{K''} \Pc \, \rck^{-\etad}
\end{align}
where the $K''$ cellular interferer locations conform to the points of the PPP $\Phic$ in $\mathcal{B}(0,R)$ while the $\mathcal{K}''$ D2D interferer locations are difficult to model in general due to the asymmetry of the voids in $\mathcal{B}(0,R)$ caused by exclusion regions. We shall turn to this issue later in the section.

\subsection{SIR Distributions}

The local-average SIRs of the cellular uplink and the D2D links are 
\begin{align}
\rhoc 
&=  \frac{\aco^{-\etac}}{\mu \sum_{j=1}^{\mathcal{K}'}\adj^{-\etac}+\frac{2\left(\mu p K+1\right)}{\etac-2}} \\
\rhod &=  \frac{ \left(\frac{\da}{K^\upbeta}\right)^{-\etad}}{\sum_{j=1}^{\mathcal{K}''}\adj^{-\etad} + \frac{1}{\mu} \sum_{k=1}^{K''} \ack^{-\etad} +\frac{2\left(p K + 1/\mu \right)}{\etad-2}}.
\end{align}
 


Given the values of $\rhoc$ and $\rhod$, i.e., conditioning on the locations within $\mathcal{B}(0,R)$, the instantaneous SIRs become exponentially distributed as in Section \ref{sec:Intf model}. 

\begin{exmp}
\label{valid_exclusion}

Consider a cellular uplink where underlaid D2D transmitters cannot occupy the circular exclusion regions. Let the normalized uplink distance be $\aco = 0.6$ while $K=10$, $\mu = 0.1$ and $\etac = 3.5$. 
The D2D interferers within the annulus $\mathcal{A}(\dc,R)$ are placed at the normalized distances $\adj=\frac{\Gamma(0.5+j)}{\sqrt{K}\Gamma(j)}$, if $\adj>\adc$, for $j=1,\ldots,10$. Shown in Fig.~\ref{Fig:Exm11} is the comparison of $F_{\SIRc | \rhoc}(\cdot)$ against the corresponding numerically computed CDF of instantaneous SIR, for different values of $\adc$. The numerical results correspond to regularly spaced circular exclusion regions within which the D2D interferers are not present (cf. Fig.~\ref{Fig:exclngrid}). Satisfactory agreement is observed for exclusion regions of various sizes.

\begin{figure}[h]  
\centering
 \includegraphics [width=0.97\columnwidth]{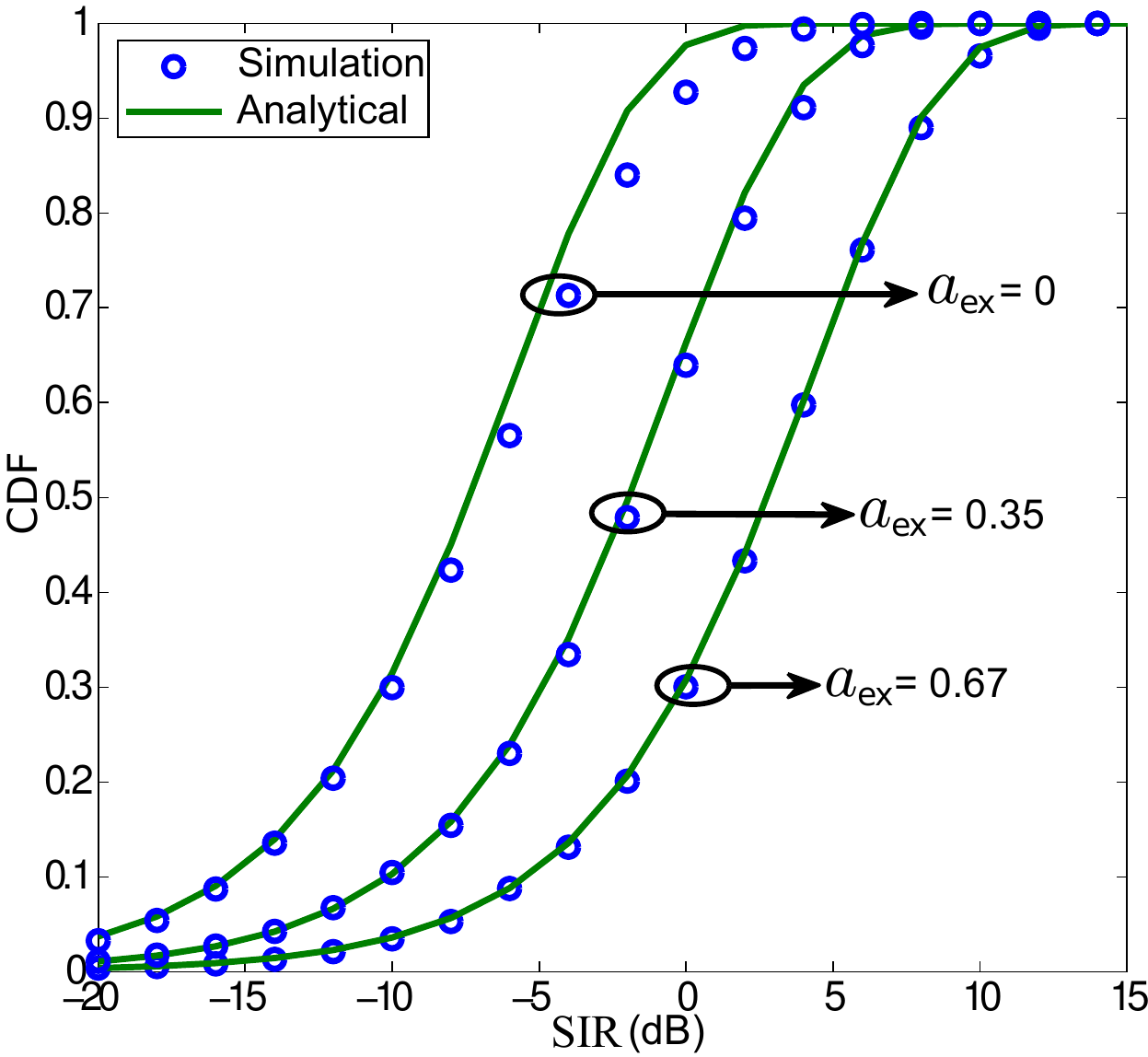}
 \caption{CDF of instantaneous SIR for a cellular uplink with underlay, $\aco = 0.6$, $\etac = 3.5$ and $K=10$.}
  \label{Fig:Exm11}
\end{figure}
\end{exmp}

\subsection{Link Spectral Efficiency}

For specific network geometries, i.e., given $\rhoc$ and $\rhod$, the link spectral efficiencies $\SEc(\rhoc)$ and $\SEd(\rhod)$ are obtained as in Section \ref{sec:Seff} and not repeated here for the sake of brevity. Then, those expressions can be further expected over the locations of the interferers inside the averaging circle, leading to the results that follow. 




\begin{prp}
\label{prp_avgcellSE_excln1}
In an underlay system with normalized exclusion regions of radius $\adc$, the uplink spectral efficiency averaged over all geometries is
\begin{align}
\label{eq:cosmos1}\nonumber
\bSEc &=  \frac{2}{e^{p K}} \int_{0}^{\infty} \frac{\log_2 e}{\gamma+1} \int_{0}^{1} a \, e^{-\gamma \, a^{\etac} \, \frac{2\left(\mu p K+1\right)}{\etac-2}} \\ 
&\quad \cdot e^{-\frac{2K}{\etac} \left[\adc^2 E_{\frac{2+\etac}{\etac}}\left(\gamma \, \mu \, \frac{a^{\etac}}{\adc^{\etac}} \right)- E_{\frac{2+\etac}{\etac}}\left(\gamma \, \mu \, a^{\etac}\right)\right]} {\rm d}a \, {\rm d}\gamma.
\end{align}
where $E_n(x)= \int_{1}^{\infty}\frac{e^{-xt}}{t^n} {\rm d}t$ is an exponential integral and $p = 1-\adc^2$.
\begin{proof}
See Appendix~\ref{App6}.
\end{proof}
\end{prp}

Next, we turn our attention to the D2D links.
Since the asymmetry of the voids present inside $\mathcal{B}(0,R)$ makes it difficult to model the D2D interferer locations corresponding to $\Iind$, we upper-bound the interference power in order to obtain a lower bound on the average spectral efficiency.
Specifically, we fill the voids inside $\mathcal{B}(0,R)$ and regard the D2D interferers as conforming to a PPP with density $\lambdad$ within $\mathcal{B}(0,R)$, which can only increase the amount of interference.

\begin{prp}
\label{prp_avgcellSE_excln2}
In an underlay system with normalized exclusion regions of radius $\adc$ and with a given $\ado$,
\begin{align}
\label{eq:cosmos2}\nonumber
\bSEd &\geq \frac{1}{e^{ K + 1} } \int_{0}^{\infty} \frac{\log_2 e}{\gamma+1} \\
&\cdot e^{-\gamma \, \frac{\da^{\etad}}{K^{\etad\upbeta}} \frac{2\left(p K + 1/\mu \right)}{\etad-2}  + \frac{2}{\etad}  \left[K  \, E_{\frac{2+\etad}{\etad}}\left(\frac{\gamma \, \da^{\etad}}{K^{\etad \upbeta}} \right) +  E_{\frac{2+\etad}{\etad}}\left(\frac{\gamma \, \da^{\etad}}{\mu  K^{\etad \upbeta}} \right)\right] } \, {\rm d}\gamma.
\end{align}
\begin{proof}
See Appendix~\ref{App7}.
\end{proof}
\end{prp} 

\begin{exmp}
\label{exclusion_valid2}

With exclusion regions around the BSs, the average system spectral efficiency (bits/s/Hz per cell) of the underlaid D2D links becomes $pK\bSEd$. In Fig.~\ref{Fig:Exm111}, the analytical lower-bound on such average system spectral efficiency is contrasted against the exact results obtained numerically for $\da = 0.12$, $\upbeta = 0$, $\mu = 0.1$ and $\etad = 4.5$.

\begin{figure}[h]  
\centering
 \includegraphics  [width=0.97\columnwidth]{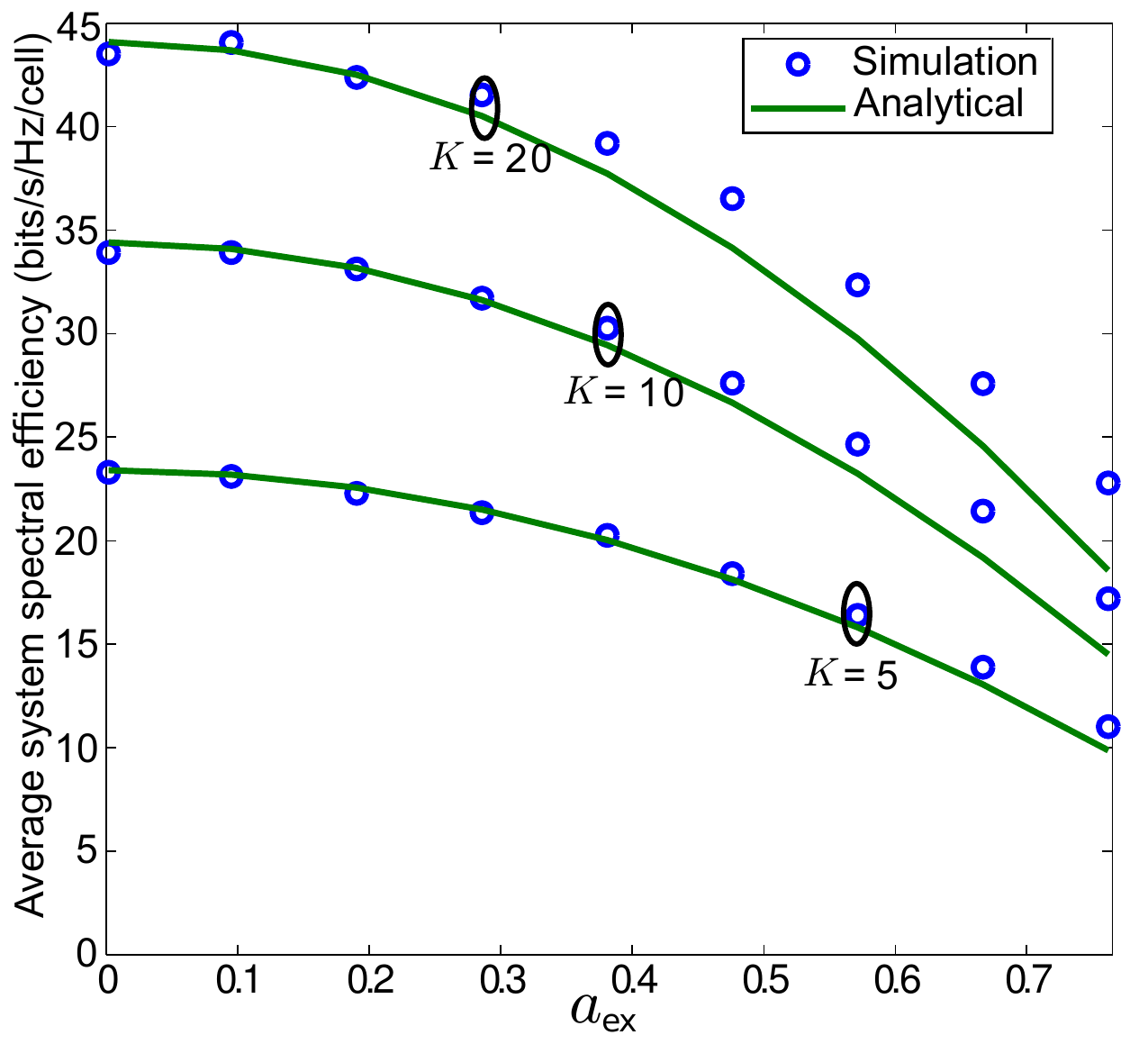}
 \caption{Average system spectral efficiency per cell of the underlaying D2D links. The analytical lower bound and exact simulation results are contrasted for different values of $K$.}
  \label{Fig:Exm111}
\end{figure}
\end{exmp}

Our final example aims at illustrating the effectiveness of the exclusion regions.

\begin{exmp}
\label{exclusion_benefits2}



With a constraint of less than 20\% degradation for the average uplink spectral efficiency, parametrized by $\upnu = 0.8$ in the relationship $\bSEc \geq \upnu \bSEc|_{K=0}$, the achievable average system spectral efficiency (bits/s/Hz per cell) of the underlaid D2D links $pK\bSEd$ computed utilizing the analytical lower-bound in (\ref{eq:cosmos2}) is plotted in Fig.~\ref{Fig:Exm17}, against the number of underlaid D2D links per cell, $pK$, for different values of $\adc$ and $\mu$. As in Example~\ref{example_underlay2}, we set $\da= 0.12$, $\upbeta = 0$, $\etac = 3.5$ and $\etad = 4.5$.
The exclusion regions allow packing more D2D links per cell, for a given degradation of the average uplink performance, thereby achieving a higher system spectral efficiency.

\begin{figure}[h]  
\centering
 \includegraphics  [width=0.97\columnwidth]{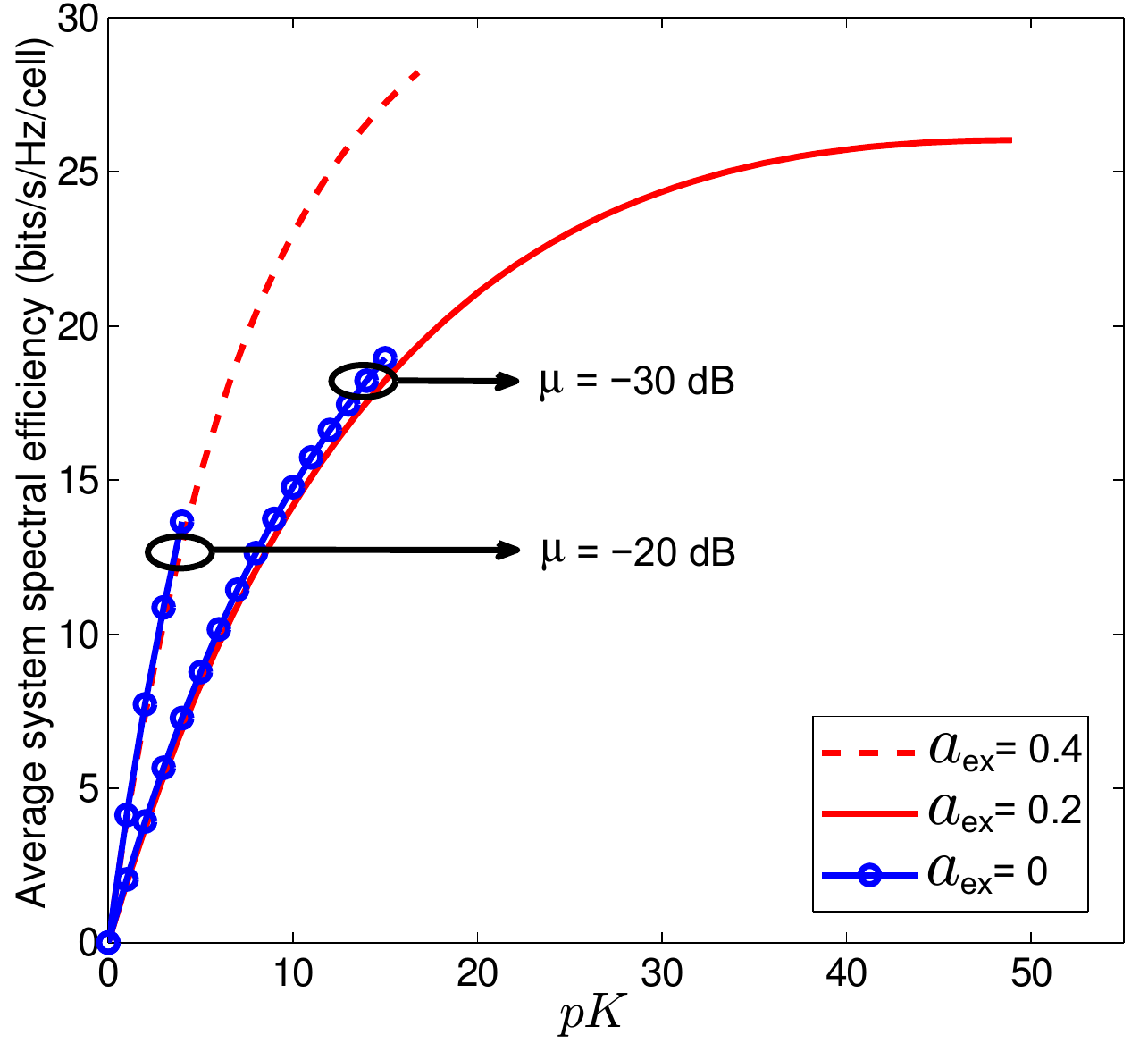}
 \caption{With $\upnu = 0.8$, achievable average system spectral efficiency of the underlaid D2D links with $\da = 0.12$ and $\upbeta = 0$, parameterized by $\mu$ and $\adc$.}
  \label{Fig:Exm17}
\end{figure}
\end{exmp}

\section{Summary}

The framework introduced in this paper enables analytical characterizations of the spectral efficiency 
of specific network geometries in addition to the average thereof, for both underlay and overlay,
 and it yields easy-to-evaluate expressions that in some cases are even in closed form.
From these expressions, and the various supporting examples, we can distill the following answers to the questions  posed in the introduction:
\begin{itemize}
\item For local traffic, direct D2D is better than uplink-downlink communication in a vast majority of situations, upwards of $80\%$ for relatively long D2D links (15\% of cell radius) and upwards of $98\%$ for shorter D2D links ($5\%$ of cell radius).
\item Tens of D2D links can be packed on each cell with acceptable mutual interference and, given a properly sized exclusion region, with only a minor effect (order $10$--$20\%$) on the cellular uplink in that cell.
\item For local traffic, the average system spectral efficiency with D2D can be between one and three orders of magnitude larger than without D2D. 
\end{itemize}

Altogether then, D2D communication offers a prime opportunity for network densification in the face of local traffic. The increase in system spectral efficiency that it can bring about is very high, even if no attempt is made to optimize the scheduling of D2D transmissions. Strong interference does arise as a problem for a share of the users, and smart scheduling can alleviate this issue; we have seen a glimpse of that by simply introducing fixed exclusions regions around the BSs.
More sophisticated schemes where the exclusion regions are dynamic could be even better and apply also to D2D links, where without smart scheduling arbitrary proximity is possible \cite{D2D-exclusion-WiOpt}.
In that sense, the extension of the analytical framework to encompass schemes such as FlashLinQ  \cite{FlashLinQ-2013} or ITLinQ \cite{ITLinQ-2014,ITLinQ-asilomar,ITLinQ-ICC} is a natural follow-up.

\section*{Acknowledgment}

The efficient editorial handling by Prof. Xiangwei Zhou and the excellent feedback provided by the reviewers are gratefully acknowledged.

\appendices
\section{Proof of Proposition~\ref{prp_CDFrhoc}}\label{Appp0}
Recall, from~(\ref{eq:SIR2}), that the local-average SIR in the presence of an averaging circle is
\begin{align}
	\rhoc &= \frac{\Pc \, \rco^{-\etac}}{\Iinc+\bIoutc}
\end{align}
where $\bIoutc$ is the spatial average of $\Ioutc$.
For this computation though, it is more convenient to retain an averaging circle (equal to a cell size) only for the cellular interferers, while not applying it to D2D interferers (or, equivalently, taking its size to infinity, which can only sharpen the model). 
With it, the local-average SIR with underlay ($\alpha=1$) becomes
\begin{align}\label{eq:rhoc_noavcr}
\rhoc &=  \frac{\rco^{-\etac}}{\mu \sum_{j=1}^{\infty}\rdj^{-\etac} + \frac{2}{(\etac-2)R^{\etac}}} . 
\end{align}


Conditioned on the summation in the denominator of (\ref{eq:rhoc_noavcr}), denoted hereafter by $\mathcal{Y} =   \sum_{j=1}^{\infty}\rdj^{-\etac}$, the CDF of $\rhoc$ can be expressed as
\begin{align}
	F_{\rhoc \,| \,\mathcal{Y}}(x) &= \mathbb{P}\left[\frac{\rco^{-\etac}}{ \mu \, \mathcal{Y} + \frac{2}{(\etac-2)R^{\etac}}} < x\right] \\
	\label{eq:cond_cdfthoc}
	&= \mathbb{P}\left[\aco > x^{-\frac{1}{\etac}} \left(\mu \, \mathcal{Y}  R^{\etac} + \frac{2}{\etac-2}\right)^{-\frac{1}{\etac}}\right].
\end{align}
For a user uniformly located in the cell, we can write the CDF of $\aco$ as
\begin{align} \label{eq:CDFrco}
F_{\aco}(a) = \left\{ \begin{array}{l l}
\aco^2 \quad & 0 \leq a \leq 1 \\
1 \quad & a> 1
\end{array} \right.
\end{align}
and thus (\ref{eq:cond_cdfthoc}) becomes
\begin{align}
F_{\rhoc \,| \,\mathcal{Y}}(x) &= 1-F_{\aco}\left[x^{-\frac{1}{\etac}} \left(\mu \, \mathcal{Y}  R^{\etac} + \frac{2}{\etac-2}\right)^{-\frac{1}{\etac}}\right] \\
\nonumber
&=1- x^{-\frac{2}{\etac}} \left(\mu \, \mathcal{Y}  R^{\etac} + \frac{2}{\etac-2}\right)^{-\frac{2}{\etac}} \\
&\qquad \qquad \qquad \qquad x \geq \frac{\etac-2}{\mu \,\mathcal{Y} R^{\etac} (\etac-2) + 2 }.
\end{align}
Expecting $F_{\rhoc \,| \,\mathcal{Y}}(x)$ over $\mathcal{Y}$, the unconditional CDF of $\rhoc$ can be obtained. Since the resulting form is unwieldy for general $\etac$, we restrict ourselves to $\etac=4$ for which the PDF of $\mathcal{Y}$ is~\cite[Sec. V]{Sousa90}
\begin{align}\label{eq:pdf_I2}
f_{\mathcal{Y}}(\nu) = \frac{\pi}{2} \, \lambdad \, \nu^{\frac{-3}{2}} e^{\frac{-\pi^3 \lambdad^2}{4\nu}} \qquad \nu>0.
\end{align}
Averaging $F_{\rhoc \,| \,\mathcal{Y}}(\cdot)$ over $\mathcal{Y}$, we obtain the CDF of $\rhoc$
\begin{align}\label{eq:CDFrhoc1}
	F_{\rhoc}(x) 
			&= \left\{ \begin{array}{l l}
			\int\limits_{\frac{1-x}{R^4 \mu x}}^{\infty} \left(	f_{\mathcal{Y}}(\nu) -  \frac{f_{\mathcal{Y}}(\nu)}{\sqrt{x \left(\mu \, \nu  R^4 + 1\right) }} \right) \, {\rm d}\nu \quad &  0<x < 1 \\ 
				1 - \int\limits_{0}^{\infty}  \frac{f_{\mathcal{Y}}(\nu)}{\sqrt{x \left(\mu \, \nu  R^4 + 1\right) }} \, {\rm d}\nu \quad &  x \geq 1  
				\end{array} \right. 				
\end{align}
where the limits of the integrals follow from $\nu \geq \frac{1-x}{R^4 \mu x}$ and $\nu > 0$. The claim of the proposition follows from inserting (\ref{eq:pdf_I2}) into (\ref{eq:CDFrhoc1}), substituting $\lambdad = \frac{K}{\pi R^2}$ and solving the integrals.
\section{Proof of Proposition~\ref{prp_CDFrhoc2}}\label{Appp00}
With overlay ($\alpha=0$), the cellular link is not subject to D2D interference and the local-average SIR 
$
\rhoc = \frac{\etac-2}{2\,\aco^{\etac}}
$
depends  only  on $\aco$. Thus, the CDF of $\rhoc$ can be expressed as
\begin{align}\label{eq:cdfrhoc3}
F_{\rhoc}(x) 
&= 1 - F_{\aco}\left[\left(\frac{\etac-2}{2\,x}\right)^{\frac{1}{\etac}}\right].
\end{align}
Applying (\ref{eq:CDFrco}) to (\ref{eq:cdfrhoc3}) yields the claimed result.

\section{Proof of Proposition~\ref{prp_CDFrhod}}\label{App5}

The distribution of $\rhod$ can be computed over the spatial locations of all interferers in the network, which is tantamount to taking the size of the averaging circle to infinity. Thereby replacing the average interference power $\bIoutd$ with $\Ioutd$ in the definition of $\rhod$ in (\ref{eq:SIR3}),
 \begin{align}\label{eq:rhod_noavcr}
	\rhod 
	&= \frac{ \rdo^{-\etad}}{ \sum_{j = 1}^{\infty}\rdj^{-\etad}+ \frac{\alpha}{\mu} \sum_{k=1}^{\infty}\rck^{-\etad}}.
 \end{align}
The denominator of (\ref{eq:rhod_noavcr}), denoted hereafter by 
$\mathcal{I}= \sum_{j = 1}^{\infty}\rdj^{-\etad}+ \frac{\alpha}{\mu} \sum_{k=1}^{\infty}\rck^{-\etad}$, 
has the characteristic function
\begin{align}\label{eq:charfun1}
\phi_{\mathcal{I}}(\omega) &= \mathbb{E}\left[e^{\mathsf{j} \, \omega \, \mathcal{I}}\right] \\
 \label{eq:charfun2}
  &= e^{-\pi \left(\lambdad + \frac{\alpha}{\mu^{2/\etad}} \lambdab \right) \Gamma\left(1-\frac{2}{\etad}\right)\,e^{-\frac{\mathsf{j}\pi}{\etad}}\, \omega^{\frac{2}{\etad}} } \quad \omega \geq 0
\end{align}
where $\mathsf{j}$ is the imaginary unit and $\phi_{\mathcal{I}}(\omega) = \phi^*_{\mathcal{I}}(-\omega)$. The expression for $\phi_{\mathcal{I}}(\omega) $ in (\ref{eq:charfun2}) is obtained as the product of the characteristic functions of the first and second summations in $\mathcal{I}$, which are computed as illustrated in~\cite[Sec. V]{Sousa90}.
Then, the density of $\mathcal{I}$ can be obtained by taking the inverse Fourier transform of $\phi_{\mathcal{I}}(\omega)$ and the corresponding CDF is
\begin{align}\label{eq:cdf_I1}\nonumber
F_{\mathcal{I}}(\nu) 
&= 1 - \frac{1}{\pi}\sum_{k=1}^{\infty} \left[\pi \left(\lambdad + \frac{\alpha}{\mu^{2/\etad}} \lambdab \right) \Gamma\left(1-\frac{2}{\etad}\right) \nu^{\frac{-2}{\etad}}\right]^k \\
&\qquad \qquad \qquad \quad \cdot \frac{\Gamma\left(\frac{2 \, k}{\etad}\right)}{k!} \sin \left[k\pi\left(1-\frac{2}{\etad}\right)\right]
\end{align}
which, for $\etad = 4$, equals
\begin{align}\label{eq:cdf_I2}
F_{\mathcal{I}}(\nu) 
= 1-\erf\left[\frac{\pi^{3/2}\left(\lambdad + \frac{\alpha}{\sqrt{\mu}} \lambdab \right)}{2 \sqrt{\nu}}\right].
\end{align}
The CDF of $\rhod$ is given by
\begin{align}\label{eq:cdf_rhod}
F_{\rhod}(x) 
			 &= 1 - F_{\mathcal{I}}\left(\frac{\rdo^{-\etad}}{x}\right).
\end{align}
Using (\ref{eq:cdf_I1}) and (\ref{eq:cdf_I2}) in (\ref{eq:cdf_rhod}), and further substituting $\rdo = \frac{\dr}{K^{\upbeta}}$ and $\lambdad = \frac{K}{\pi R^2}$, we obtain the claim of Prop.~\ref{prp_CDFrhod}.

\section{Proof of Proposition~\ref{prp_avg_cellSE3}}\label{App4}


The uplink spectral efficiency averaged over all geometries is
	 	$
		\bSEc = \mathbb{E}[\SEc(\rhoc)] 
		$
	 	with expectation over $\rho$. Expanding this expectation,
	 	\begin{align}
	 	\label{eq:cellSEproof1}
	    \bSEc &= \mathbb{E}\left[\mathbb{E}\left[ \log_2(1+\SIRc|\rhoc) \right] \right] \\
	 	\label{eq:cellSEproof2}
	 	&= \mathbb{E}\left[  \int_{0}^{\infty} \frac{\log_2 e}{\gamma+1}(1-F_{\SIRc|\rhoc}(\gamma)) \, {\rm d}\gamma \right] \\
	 	&=  \int_{0}^{\infty} \frac{\log_2 e}{\gamma+1}\left(1- \mathbb{E}\left[F_{\SIRc|\rhoc}(\gamma) \right]\right) \, {\rm d}\gamma \\
	 	\label{eq:cellSEproof3}
	 	&= \int_{0}^{\infty} \frac{\log_2 e}{\gamma+1}\left(1- F_{\SIRc}(\gamma) \right) \, {\rm d}\gamma 	 	
	 	\end{align} 
where the outer and inner expectations in (\ref{eq:cellSEproof1}) are over $\rhoc$ and over the fading, respectively. We next compute $ F_{\SIRc}(\cdot)$ and then use it to evaluate~(\ref{eq:cellSEproof3}).
For this computation, we employ the approach in Appendix~\ref{Appp0} and utilize (\ref{eq:rhoc_noavcr}) to expand the conditional CDF of the SIR
\begin{align}
F_{\SIRc | \rhoc}(\gamma) 
	&= 1 - e^{-\gamma \, \frac{\rco^{\etac}}{R^{\etac}} \frac{2}{\etac-2}} \prod_{j=1}^{\infty} e^{-\gamma \,\alpha\, \mu \, \rco^{\etac} \, \rdj^{-\etac}}. 
\end{align}
Conditioning on $\rco$ and averaging over the interference locations
\begin{align}
\label{eq:eqn1}
F_{\SIRc | \rco}(\gamma) &= 1 - e^{-\gamma \frac{\rco^{\etac}}{R^{\etac}} \frac{2}{\etac-2}} \,  \mathbb{E}\! \left[ \prod_{j=1}^{\infty} e^{-\gamma \,\alpha \mu \, \rco^{\etac} \, \rdj^{-\etac}} \right]  \\
\label{eq:eqn2}\nonumber
 &= 1 - e^{-\gamma \frac{\rco^{\etac}}{R^{\etac}} \frac{2}{\etac-2}} \\
   &\qquad \quad \cdot e^{- 2 \pi \lambdad \int_{0}^{\infty} \left(1-e^{-\gamma\,\alpha \mu \, \rco^{\etac} \, x^{-\etac}} \right) x {\rm d}x}  \\
 \label{eq:eqn3}\nonumber
 &= 1 - e^{-\gamma \frac{\rco^{\etac}}{R^{\etac}} \frac{2}{\etac-2}}  \\
&\qquad \cdot e^{-\gamma^{\frac{2}{\etac}} \, \frac{\rco^2}{R^2} K \frac{2}{\etac} \int_{0}^{\infty} \left(1-e^{-\alpha \mu u} \right) \frac{1}{u^{1+2/\etac}} {\rm d}u} 
\end{align}
where the expectation in (\ref{eq:eqn1}) is over the PPP $\Phid$, (\ref{eq:eqn2}) follows from the definition of the probability generating functional (PGFL) of the PPP~\cite{MHaenggi12}, and (\ref{eq:eqn3}) follows from the variable change $\gamma \, \rco^{\etac} \, x^{-\etac} = u$ and the relation $\pi\lambdad = K/R^2$. Employing integration by parts in (\ref{eq:eqn3}) and invoking $\aco=\frac{\rco}{R}$, we obtain
\begin{align}
\label{eq:underlaycellse2}
F_{\SIRc | \aco}(\gamma) &= 1- e^{-\gamma \, \aco^{\etac} \frac{2}{\etac-2}\,-\alpha\,(\gamma\mu)^{\frac{2}{\etac}} \, \aco^2 \, K \, \Gamma\left(1-\frac{2}{\etac}\right)}.
\end{align}


Further averaging $F_{\SIRc | \aco}(\cdot)$ over $\aco$ via the distribution in (\ref{eq:CDFrco}) we obtain $F_{\SIRc}(\cdot)$, which plugged into (\ref{eq:cellSEproof3}) yields (\ref{eq:conditional_avSE}).

\section{Proof of Proposition~\ref{prp_avg_cellSE1}}\label{App3}

Again, with overlay ($\alpha=0$), the local-average SIR of the cellular uplink is
$
\rhoc = \frac{\etac-2}{2\,\aco^{\etac}}
$
and the expectation is only over $\aco$.
Averaging $\SEc(\rhoc)$ over $\aco$ via the distribution in (\ref{eq:CDFrco}) yields
\begin{align}\label{eq:avcellseproof0}
\bSEc 
=\log_2(e) \int_{0}^{1} e^{\frac{2\,a^{\etac}}{\etac-2}} \, E_1 \!\left(\frac{2\,a^{\etac}}{\etac-2}\right) {\rm d}F_{\aco}(a).
\end{align}
When $x$ is positive and real, $E_1(x)=-E_\text{i}(-x)$ where $E_\text{i}(x) = \int_{-x}^{\infty}\frac{-e^{-t}}{t} \, {\rm d}t$. Utilizing this relation in~(\ref{eq:avcellseproof0}) and then evaluating the integral by virtue of the identity given in~\cite{WOLFRAM1}, we obtain~(\ref{eq:UL seff Overlay}).

\section{Proof of Proposition~\ref{avg_D2DSE1}} \label{App1}

Here we can directly average the link spectral efficiency over the spatial locations of all cellular and D2D interferers in the network, 
without invoking any a-priori averaging circle, using (\ref{eq:rhod_noavcr}). 
Thus, the conditional CDF of $\SIRd$ is expanded as
\begin{align} 
F_{\SIRd|\rhod}(\gamma)
&= 1-\prod_{j=1}^{\infty} e^{-\gamma \, \rdo^{\etad} \, \rdj^{-\etad}} \prod_{k=1}^{\infty} e^{-\gamma \, \frac{\alpha}{\mu} \, \rdo^{\etad} \, \rck^{-\etad}} . \label{eq:SIRcond}
\end{align}

Maintaining the conditioning on the desired link distance $\rdo$, we average (\ref{eq:SIRcond}) over the PPPs $\Phid$ and $\Phic$ to obtain
\begin{align} 
F_{\SIRd|\ado}(\gamma) 
&= 1-\mathbb{E}_{\Phid}\!\!\left[\prod_{j = 1}^{\infty} e^{-\gamma \, \rdo^{\etad} \, \rdj^{-\etad}  }\right] \nonumber \\
&\qquad \cdot \mathbb{E}_{\Phic}\!\!\left[\prod_{k = 1}^{\infty} e^{-\gamma \, \frac{\alpha}{\mu}  \, \rdo^{\etad} \, \rck^{-\etad}  }\right] \label{eq:cdfsira2b}\\
&=  1-e^{-\gamma^{\frac{2}{\etad}}  \frac{\da^2}{K^{2\upbeta}}  \Gamma\left(1-\frac{2}{\etad}\right) \, \left(K+ \alpha \, \mu^{\frac{-2}{\etad}} \right)} \label{eq:cdfsirc3} \end{align}
where 
the expectations in (\ref{eq:cdfsirc3}) are computed  as in Appendix~\ref{App4} with the substitution $\ado=\frac{\da}{K^\upbeta}$.
Thus, the average spectral efficiency of the D2D links becomes
\begin{align}\label{eq:sedd}
\bSEd &= \int_{0}^{\infty} \frac{\log_2 e}{\gamma+1}(1-F_{\SIRd|\ado}(\gamma)) \, {\rm d}\gamma \\
	&= \log_2(e)  \int_{0}^{\infty} \frac{e^{-\gamma^{\frac{2}{\etad}} \frac{\da^2}{K^{2\upbeta}} \Gamma\left(1-\frac{2}{\etad}\right) \, \left(K+ \frac{\alpha}{\mu^{2/\etad}}\right)} }{\gamma+1} \, {\rm d}\gamma \label{eq:sedd2}
\end{align}
which is unwieldy for general $\etad$. However, for $\etad = 4$,~(\ref{eq:sedd2}) reduces to 
\begin{align}
\bSEd 
&= 2\log_2(e)\int_{0}^{\infty} \frac{x\,e^{-x \, \mathcal{K} \da^2}}{x^2+1} \,  {\rm d}x 
\label{eq:sed3ad} 
\end{align}
which follows from the variable change $\sqrt{\gamma}=x$ in~(\ref{eq:sedd}) with $\mathcal{K}=\frac{\sqrt{\pi}}{K^{2\upbeta}}(K+\frac{ \alpha}{\sqrt{\mu}})$. By virtue of~\cite[3.354.2]{Gradshteyn-07},~(\ref{eq:sed3ad}) turns into the claimed expression in~(\ref{eq:Seff D2D eta4}).

\section{Proof of Proposition~\ref{prp_avgcellSE_excln1}}\label{App6}

The conditional CDF of $\SIRc$ is
\begin{align}
F_{\SIRc | \rhoc}(\gamma) 
	&= 1- e^{- \gamma \, \rco^{\etac} \, \bIoutc}  \prod_{j=1}^{\mathcal{K}'} e^{-\gamma \,\mu \, \rco^{\etac} \, \rdj^{-\etac}}.
\end{align}
Conditioning on $\rco$ and averaging over the distances $\{\rdj\}_{j=1}^{\mathcal{K}'}$ to the D2D interferers in the annulus $\mathcal{A}(\dc,R)$, we obtain

\begin{align}
F_{\SIRc | \aco}(\gamma) &= 1- e^{- \gamma \, \rco^{\etac} \, \bIoutc} \, \mathbb{E}_{\Phid}\!\!\left[ \prod_{j=1}^{\mathcal{K}'} e^{-\gamma \,\mu \, \rco^{\etac} \, \rdj^{-\etac}} \right] \\
\label{eq:excln_proof_c1}
&= 1- e^{- \gamma \, \rco^{\etac} \, \bIoutc \,-\, 2\pi\lambdad \int_{\dc}^{R} \left(1-e^{-\gamma \,\mu \, \rco^{\etac} \, x^{-\etac}}\right) x \, {\rm d}x} \\
\label{eq:excln_proof_c2}
&= 1-e^{-p K - \gamma \, \aco^{\etac} \, \frac{2\left(\mu p K+1\right)}{\etac-2} } \nonumber \\ 
&\qquad \cdot e^{- \frac{2K}{\etac} \left[\adc^2 E_{\frac{2+\etac}{\etac}}\left(\gamma \, \mu \, \frac{\aco^{\etac}}{\adc^{\etac}} \right) \,-\, E_{\frac{2+\etac}{\etac}}\left(\gamma \, \mu \, \aco^{\etac}\right)\right]} 
\end{align}
where (\ref{eq:excln_proof_c1}) follows from the definition of the PGFL of PPP, the integral is solved following the approach in Appendix~\ref{App4}, and we substitute $\frac{\rco}{R}=\aco$ to obtain (\ref{eq:excln_proof_c2}). 
Eq. (\ref{eq:excln_proof_c2}) is further averaged over $\aco$ via (\ref{eq:CDFrco}) to obtain $F_{\SIRc}(\cdot)$, which plugged into (\ref{eq:cellSEproof3}) yields the claimed result in (\ref{eq:cosmos1}).

\section{Proof of Proposition~\ref{prp_avgcellSE_excln2}}\label{App7}

The conditional CDF of $\SIRd$ is

\begin{align}
F_{\SIRd | \ado}(\gamma) &= 1- e^{- \gamma \, \rdo^{\etad} \, \bIoutd} \nonumber \\ 
&\quad  \cdot \mathbb{E}_{\Phid}\!\!\left[  \prod_{j=1}^{\mathcal{K}''} e^{-\gamma \, \rdo^{\etad} \, \rdj^{-\etad}} \right] \, \mathbb{E}_{\Phic}\!\!\left[ \prod_{k=1}^{K''} e^{-\frac{\gamma}{\mu} \rdo^{\etad} \, \rck^{-\etad}} \right]  \\
&= 1- e^{- \gamma \, \rdo^{\etad} \, \bIoutd} e^{-2\pi\lambdad \int_{0}^{R} \left(1-e^{-\gamma \, \rdo^{\etad} \, x^{-\etad}}\right) x \, {\rm d}x }  \nonumber \\
&\qquad \cdot  e^{-2\pi\lambdab \int_{0}^{R} \left(1-e^{-\frac{\gamma}{\mu}  \rdo^{\etad} \, x^{-\etad}}\right) x \, {\rm d}x } \label{eq:excln_proofd1} \\
&= 1-e^{-\gamma \, \ado^{\etad} \frac{2}{\etad-2} \left(p K + \frac{1}{\mu} \right)} e^{ -(K + 1) } \nonumber \\ 
&\qquad \cdot e^{ \frac{2}{\etad}  \left[K  \, E_{\frac{2+\etad}{\etad}}\left(\gamma \frac{\da^{\etad}}{K^{\etad\upbeta}} \right) +  E_{\frac{2+\etad}{\etad}}\left(\frac{\gamma }{\mu} \frac{\da^{\etad}}{K^{\etad\upbeta}} \right)\right]} \label{eq:excln_proof_d2}
\end{align}
where (\ref{eq:excln_proofd1}) follows from the definition of the PGFL of PPP, the integrals are solved as in Appendix \ref{App4} and we substitute $\ado = \frac{\da}{K^\upbeta}$ to obtain (\ref{eq:excln_proof_d2}). Then, (\ref{eq:cosmos2}) follows from plugging (\ref{eq:excln_proof_d2}) into (\ref{eq:sedd}).

\bibliographystyle{IEEEbib}
\bibliography{jour_short,conf_short,refs_Ratheesh}

\end{document}